\documentclass[12pt,preprint]{aastex}

\def\ae2{{\it Astro-E2}}

\newcommand{\etal}{{\it et al.} }

\def\km/s{${rm km\ s^{-1}}$}

\begin{document}

\title{Simultaneous Ultraviolet and X-ray Observations of the Seyfert Galaxy
NGC 4151. I. Physical Conditions in the X-ray Absorbers}

\author{S. B. Kraemer\altaffilmark{1,2},
I.M. George\altaffilmark{3,4},
D.M. Crenshaw\altaffilmark{5},
J.R. Gabel\altaffilmark{6},
T.J. Turner\altaffilmark{3,4},
T.R. Gull\altaffilmark{2},
J.B. Hutchings\altaffilmark{7},
G.A. Kriss\altaffilmark{8},
R.F. Mushotzky\altaffilmark{4},
H. Netzer\altaffilmark{9},
B.M. Peterson\altaffilmark{10},
\& Ehud Behar\altaffilmark{11}}

\altaffiltext{1}{Institute for Astrophysics and Computational Sciences, 
Department of Physics , The Catholic University of America,
Washington, DC 20064; kraemer@yancey.gsfc.nasa.gov}

\altaffiltext{2}{Exploration of the Universe Division, Code 667, NASA's Goddard Space Flight 
Center, Greenbelt, MD  20771}

\altaffiltext{3}{Physics Department, University of Maryland, Baltimore
County, Baltimore, MD 21250}

\altaffiltext{4}{Exploration of the Universe Division, Code 662, 
NASA Goddard Space Flight Center, Greenbelt, MD 20771}

\altaffiltext{5}{Department of Physics and Astronomy, Georgia State University,
Atlanta, GA 30303}

\altaffiltext{6}{University of Colorado, CASA, UCB 389, Boulder, CO 80309-0389}

\altaffiltext{7}{Dominion Astrophysical Observatory, National Research Council of Canada, 
Victoria, BC V8X4M6, Canada}

\altaffiltext{8}{Space Telescope Science Institute, 3800 San Martin Dr. Baltimore, Md 21218}

\altaffiltext{9}{School of Physics and Astronomy, Raymond and Beverly Sackler Faculty of
Exact Sciences, Tel-Aviv University, Tel-Aviv 69978, Israel}

\altaffiltext{10}{Department of Astronomy, The Ohio State University, 140 W. 18th Ave.,
Columbus, OH 43210-106}

\altaffiltext{11}{Department of Physics, Technion, Haifa, 32000, Israel} 

\begin{abstract}
We present a detailed analysis of the intrinsic X-ray absorption in
the Seyfert 1 galaxy NGC 4151 using {\it Chandra}/High Energy
Transmission Grating Spectrometer data obtained 2002 May as part of a
program which included simultaneous ultraviolet (UV) spectra using the
{\it Hubble Space Telescope}/Space Telescope Imaging Spectrograph and
the {\it Far Ultraviolet Spectrographic Explorer}. Previous studies,
most recently using {\it ASCA} spectra, revealed a large ($>$
10$^{22}$ cm$^{-2}$) column of intervening gas, which has varied both
in ionization state and total column density. NGC 4151 was in a
relatively low flux state during 
the observations reported here 
($\sim$ 25\% of
its historic maximum), although roughly 2.5 times as bright in the 2
--10 keV band as during a {\it Chandra} observation in 2000.  At both
epochs, the soft X-ray band was dominated by emission lines, which
show no discernible variation in flux between the two
observations. The 2002 {\it Chandra} data show the presence of a very
highly ionized absorber, in the form of H-like and He-like Mg, Si, and
S lines, as well as lower ionization gas  
via the presence of inner-shell
absorption lines from lower-ionization species of these elements.
The latter accounts for
both the bulk of the soft X-ray absorption and the high covering
factor UV absorption lines of O~VI, C~IV, and N~V with outflow velocities 
$\approx$ 500 km
s$^{-1}$. The presence of high ionization gas, which is not easily
detected at low resolution (e.g., with {\it ASCA}), appears common
among Seyfert galaxies. Since this gas is too highly ionized to be
radiatively accelerated in sources such as NGC 4151, which  
is radiating
at a small fraction of its Eddington Luminosity, it may be key to
understanding the dynamics of mass outflow.  We find that the deeper
broad-band absorption detected in the 2000 {\it Chandra} data is the
result of both 1) lower ionization of the intervening gas due 
to the lower ionizing flux and 2) a factor of $\sim$ 3 higher
column density of the lower ionization component. To account for
this bulk motion, we estimate that this component must have a velocity
$\gtrsim$ 1250 km s$^{-1}$ transverse to our line-of-sight. This is
consistent with the rotational velocity of gas arising from the
putative accretion disk. While both thermal wind and
magneto-hydrodynamic models predict large non-radial motions, we
suggest that the latter mechanism is more consistent with the results
of the photoionization models of the absorbers

\end{abstract}


\section{Introduction}
\label{Sec:Intro}

Mass outflow, revealed by intrinsic blue-shifted absorption lines in
ultraviolet (UV) and X-ray spectra, appears to be ubiquitous among AGN
(Crenshaw, Kraemer, \& George 2003). For example, more than 50\% of
Seyfert galaxies observed with the spectrographs aboard the {\it
Hubble Space Telescope (HST)} (Crenshaw et al. 1999) and with the {\it
Far Ultraviolet Spectroscopic Explorer (FUSE)} (Kriss 2002) show UV
absorption lines. While X-ray spectra obtained with {\it ASCA} showed
absorption features such as bound-free edges of O~VII and
O~VIII, in a similarly large fraction of Seyferts (Reynolds 1997:
George et al. 1998), the blue-shifted absorption lines detected in
high-resolution spectra from {\it Chandra} and {\it XMM-Newton} proved
that the X-ray absorbing gas is also outflowing from the nuclei of
these galaxies (e.g. Kaastra et al. 2000, 2002; Kaspi et al. 2000a;
2001; 2002).  The high global covering factors and kinetic luminosities
inferred for the intrinsic absorbers suggests that they are a critical
component of the central regions of AGN (Crenshaw et al. 2003, and
references  
therein). However, many open questions remain, such as the physical
connection, if any, between the UV and X-ray absorbers (e.g. Mathur et
al. 1994; Kriss et al. 1995, 1996; Crenshaw \& Kraemer 1999; Kraemer
et al. 2002), the mechanism by which the outflowing gas is accelerated,
e.g. thermal winds (Krolik \& Begelman 1986), radiation pressure
(Murray et al. 1995), or magnetically driven winds (Bottorff et
al. 2000), and from whence the flows originate.

NGC 4151 ($cz = 995$ km s$^{-1}$; determined from 21 cm observations:
de Vaucouleurs et al. 1991), the nearest bright Seyfert 1 galaxy, has
been extensively studied at all wavelengths. Ultraviolet observations,
including those obtained with {\it IUE} (Boksenberg et al. 1978), the
Hopkins Ultraviolet Telescope (Kriss et al. 1992), the {\it
HST}/Goddard High Resolution Spectrograph (Weymann et al. 1997) and
the {\it HST}/Space Telescope Imaging Spectrograph (STIS) (Crenshaw et
al. 2000), revealed strong intrinsic absorption lines from species
covering a wide range in ionization potential, as well as those from
excited and metastable states of a number of ions. There is strong
evidence that the ionization state of the UV absorbers is correlated
with the flux of the continuum source (Bromage et al. 1985; Kraemer et
al. 2001), as expected in photo-ionized gas. X-ray spectra show the
effect of absorption by a large column of gas (Ives, Sanford, \& Penston 1976), 
which is variable (e.g. Barr et al. 1977)
and ionized (Yaqoob, Warwick, \& Pounds 1989). Based
on {\it ASCA} observations, George et al.  (1998) were able to model
the 0.6--10 keV X-ray continuum as a flat power law (photon index
$\Gamma$ $\sim$ 1.5), modified by a photo-ionized absorber of column
density $N_{H}$ $\sim$ 10$^{22-23}$ cm$^{-2}$. The gas was found to be in
a fairly high ionization state, characterized by an ionization
parameter, $U \sim 1$, where,
\begin{equation}
U = \frac{Q}{4\pi~r^{2}~c~n_{H}},
\end{equation}  
and $r$ is the radial distance of the absorber, $n_{H}$ is
the number density of hydrogen, and $Q = \int_{13.6 eV}^{\infty}\frac{L_{\nu}}{h\nu}~d\nu$,
or the number of ionizing photons s$^{-1}$ emitted by a source of luminosity $L_{\nu}$
\footnote{In George et al., the models
were parameterized in terms of the ``X-ray'' ionization parameter, 
$U_{x}$, for which the integration is taken from 0.1 keV
to 10 keV. They determined that $U_{x} = 0.06 - 0.08$; note that the conversion 
between $U$ and $U_{x}$ is sensitive to the shape of the ionizing 
continuum below 0.1 keV.}. Interestingly, the changes in the opacity
of the X-ray absorber did not seem to depend on the flux state of the
intrinsic continuum, and, in fact, there is some evidence that they
are anti-correlated (see Weaver et al. 1994), which may be evidence
for both a multi-component absorbing medium and bulk motion across our
line-of-sight.

More recently, Schurch \& Warwick (2002) used {\it BeppoSax} data from
1999 January and a series of {\it ASCA} observations from 2000 May 15
-- 20 to attempt to determine the nature of the X-ray absorber in NGC
4151. They were able to fit the data with a flat underlying power-law
($\Gamma \approx$ 1.65), a contribution from a cold reflector, and a
two-component absorber: an intrinsic neutral component with $N_{H}$
$\approx$ 3.4 x 10$^{22}$ cm$^{-2}$ and a highly ionized absorber,
with $N_{H}$ $\approx$ 2.4 x 10$^{23}$ cm$^{-2}$ and an ionization
parameter $U \sim$ 10. However, Kraemer et al. (2001) analyzed UV
spectra, obtained with STIS in 1999 July, and determined that the
intrinsic column of low-ionization gas could not exceed a few x
10$^{21}$ cm$^{-2}$, although it is possible that the amount of
low-ionization gas was extremely variable over this period. The lack
of contemporaneous high-resolution UV and X-ray spectra have hindered
attempts to unravel the various components of intrinsic absorption in
this object.  To that end, we obtained near-simultaneous STIS, {\it
FUSE}, and {\it Chandra}/High Energy Transmission Grating Spectrometer
(HETGS) observations on 2002 May 9 --10. In this paper, we present our
analysis of the X-ray continuum ($\sim$1--10~keV) and the imprinted
absorption lines.  The UV results will be presented in a subsequent
paper (S.B. Kraemer et al., in preparation; hereafter Paper II), as
will the results from a study of the soft X-ray emission lines and
reflection component (I.M. George et al. in preparation, hereafter Paper III).

\section{Observation and Data Reduction}
\label{Sec:Obs}

The main data set considered in this paper was obtained 
during a {\it Chandra} observation of NGC~4151 over the period 
2002 May 07 23:26 to May 11 14:14 (UTC). The observation
was performed in two parts ({\it Chandra} `OBSIDs' 3480 and 3052), 
with a gap of $\sim$17~hr (May 09 01:34--18:26) during 
perigee and passage of the spacecraft through the Earth's 
radiation belts. 
The HETGS
(Marshall, Dewey \& Ishibashi 2004) was deployed in the
optical path, and 
the full 
`S-array' of the Advanced CCD Imaging Spectrometer (ACIS; 
Garmire et al 2003) was used as the focal-plane detector
(with an operating temperature of 153~K).
To avoid the risk of telemetry saturation, 
only the central half of the ACIS-S array was read out, 
using a CCD frame-time of 1.8~s. The target was slightly offset 
from the nominal pointing direction (by 20~arcsec) to 
ensure the data within the zeroth-order image was always 
read-out using the same CCD node during the intentional dithering 
of the spacecraft.

In this paper we also refer to two other {\it Chandra} observations 
of NGC~4151. The first is that performed in 
2000 March (OBSID 335) with a duration of 
$\sim$50~ks, and also using the HETGS/ACIS-S instrumental 
configuration. Some of the results of this observation have been 
presented previously by Ogle \etal (2000).
The second additional {\it Chandra} observation referred to here 
was performed over the period 2002 July 02--03 (OBSID 3089).
The ACIS-S array  was also employed as the focal-plane detector for this
observation, but with the Low-Energy Transmission 
Grating Spectrometer (LETGS; Brinkman et al. 1997) deployed 
in place of the HETGS.

\subsection{Data Reduction}
\label{Sec:reduction}

The HETGS data from both epochs were processed using the 
{\tt ciao} (v3.0.1) and {\tt ftools} (v5.3.1) software 
packages, along with calibration files from the 
{\it Chandra} {\tt CALDB} (v2.26) release\footnote{Including
the {\tt acisD1999-08-13contamN0003.fits} file to correct for 
the time-dependent contamination on the surface of the CCDs
(e.g. see Marshall et al. 2004).}.
In the case of the observation using the LETGS, we used
the more recent {\tt ciao} (v3.2.0) and {\tt CALDB} (v3.0.0)
in order to incorporate the latest estimates of the 
spatial and temporal variations in this 
contamination\footnote{We have performed spot checks, and the 
use of the latest calibration files and software 
do not affect the conclusions presented here.}.
We also made use of our own
software for a number of the analysis tasks.
The data were processed in the standard manner, 
including the removal of bad detector pixels and other detector
artifacts, and removal of events with detector 
`grades' {\it not} equal to 0, 2, 3, 4, or 6.
The selection criteria used resulted in
exposures of $\sim$48~ks for the HETGS observation in 2000, 
$\sim$93 and $\sim$157~ks for the 
first and second segments of the HETGS observation in 2002 May, 
and $\sim$84~ks for the LETGS observation in 2002 July. 

The dispersed source spectra for the various grating arms
were extracted using a strip centered on the corresponding 
dispersed spectrum, with a half-width 
(in the cross-dispersion direction) of $5\times10^{-4}$ degrees.
This corresponds to a distance of $\sim$ 115 pc on the plane of the 
sky assuming a distance of 13.2~Mpc to NGC~4151. 
We are only concerned about the nucleus, although the extracted
spectrum  
will
include a contribution from the X-ray emission-line gas
in the inner Narrow-Line Region (see Ogle et al. 2000).

The intrinsic energy resolution of the CCDs were used to 
distinguish between the over-lapping orders in the dispersed 
spectra. The total number of counts in the source extraction 
cells with energies and positions consistent with the
$\pm1^{\rm st}$-order MEG spectra in the 0.5--7~keV band
were 9436 for the observation in 2000, 
and 173186 for the combined HETGS observations in 2002 May.
The corresponding number in the LEG during the observation in 
2002 July was 97448 counts.
From an analysis of strips adjacent and parallel to the source 
extraction cell, we estimate that in all cases $<2$\% of the 
counts in the extraction cell are due to 
background events
(including the extended X-ray emission surrounding the nucleus).
This background
was therefore ignored during the analysis 
of the nuclear continuum presented here.

The line-spread function of each of the gratings 
can be well-approximated by a Gaussian at all energies
with a width $\sigma \simeq 2\times10^{-2}$, $\simeq 10^{-2}$, and 
$\simeq 5\times10^{-3}$~\AA, for the 
$1^{\rm st}$-order spectra of the LEG, MEG, and HEG (respectively).
All the spectra were initially binned to this size, and the 
positive and negative orders were combined.

\subsection{Temporal Analysis}
\label{Sec:timing}

\begin{figure}
\includegraphics[scale=0.60,angle=270]{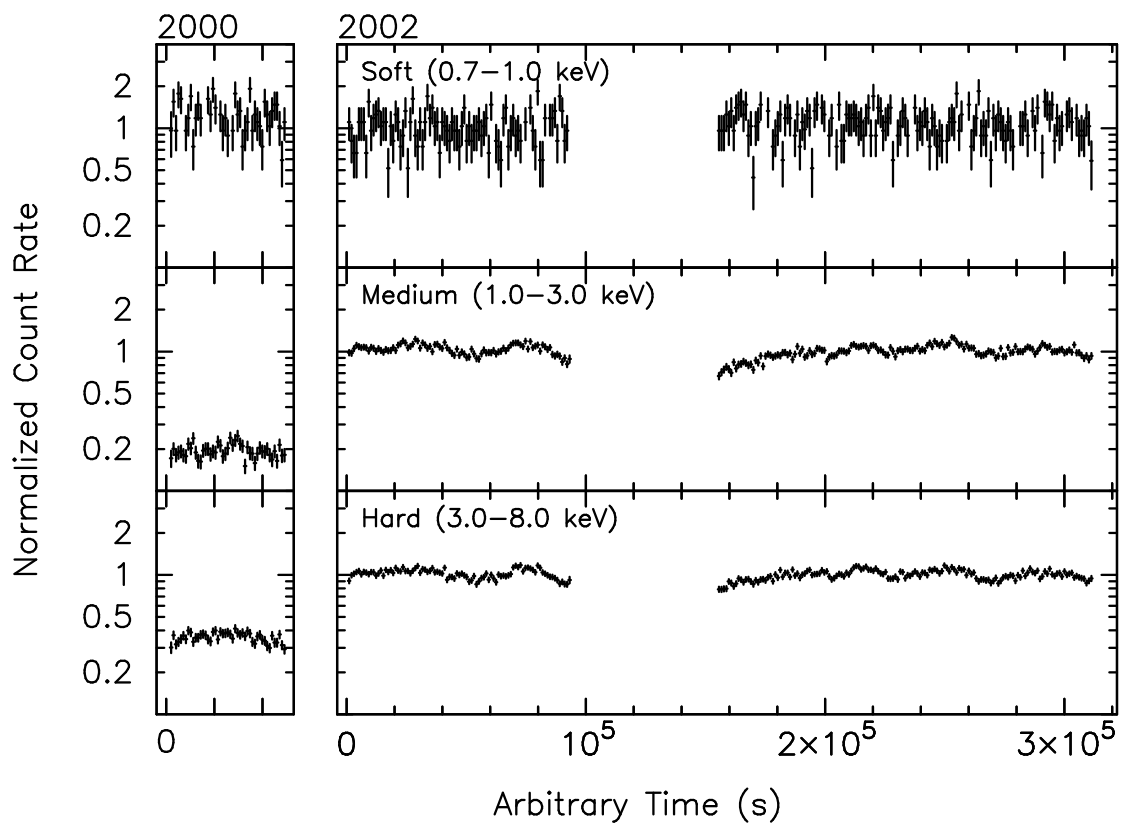}
\caption{Coadded MEG and HEG $1^{\rm st}$-order light curves, using 1024s 
bins, from the two HETGS observations discussed here. For simplicity the light
curves in each band have been renormalized to the mean count rate obtained 
during
the 2002 observations. In the soft
(0.5--1.0~keV) band, the light curves are consistent with a 
single constant flux both within each observation, and at each epoch.
Variability is apparent during both observations 
in the medium (1.0-3.0~keV) and hard (3.0-8.0~keV) bands. 
More dramatically, NGC~4151 brightened considerably in these bands 
between 2000 and 2002, with the larger fractional increase occurring 
in the medium band.
\label{fig:lc1024}}
\end{figure}

The variability exhibited by the source during and between 
the two HETGS observations is illustrated in 
Fig.~\ref{fig:lc1024}, which shows the coadded MEG and HEG 
$1^{\rm st}$-order light curves in three energy bands.
The data are shown using 1024~s bin, and 
for simplicity each light curve has been normalized to the 
error-weighted mean value of the (combined) 2002 observation.
We find the count rate in the soft X-ray band (0.7--1.0~keV)
during the 2002 observation to be consistent with a constant  
($1.32\pm0.35\times10^{-2}\ {\rm count\ s^{-1}}$; 
$\chi^{2}=213$ for 219 degrees of freedom [dof]).
The earlier 2000 observation has a consistent mean value 
($1.52\pm0.42\times10^{-2}\ {\rm count\ s^{-1}}$) in this band, 
although with a hint of variability within the observation
($\chi^{2}=51$ for 41 dof).
More interestingly, Fig.~\ref{fig:lc1024} clearly shows 
the source exhibited a dramatic change in intensity in both 
the medium and hard bands.
In the medium (1.0-3.0~keV) band, NGC~4151 brightened by a 
factor of $\sim$5 (to $0.52\pm0.05\ {\rm count\ s^{-1}}$)
between 2000 and 2002, while in the 
hard (3.0-8.0~keV) band the increase was a factor  $\sim$3 
(to $0.67\pm0.05\ {\rm count\ s^{-1}}$)
between these epochs.
The source does exhibit statistically-significant variations 
in intensity in both these bands during both epochs. 
However this is of relatively-low amplitude 
(maximum trough-to-peak amplitude of a factor $\sim 1.9$
in the medium band during the 2002 observations).

Light curves were also constructed in the above bands 
for the LETGS observation in 2002 July. The soft band 
light curve was found to be consistent with a constant  
($2.21\pm0.04\times10^{-2}\ {\rm count\ s^{-1}}$; 
$\chi^{2}=64$ for 80 dof) during this epoch. 
As will be illustrated below, correcting for the differences
in the spectral responses of the instruments, this LETGS count rate is 
consistent with the HETGS count rates  
quoted above in this band.
The mean count rates in the medium and hard X-ray bands during the 
LETGS observations were 
$0.42\pm0.04\ {\rm count\ s^{-1}}$
and 
$0.64\pm0.02\ {\rm count\ s^{-1}}$ (respectively). 
As for the HETGS data, statistically-significant variations 
were evident in both bands during this epoch, again with 
maximum trough-to-peak amplitude of a factor $\sim 1.9$.
Thus hereafter we consider only the time-averaged spectra 
obtained during each epoch.

\section{Spectral Analysis and Modeling}

\subsection{The Spectral Signature of the Intrinsic Absorber}
\label{Sec:abs}

\begin{figure}
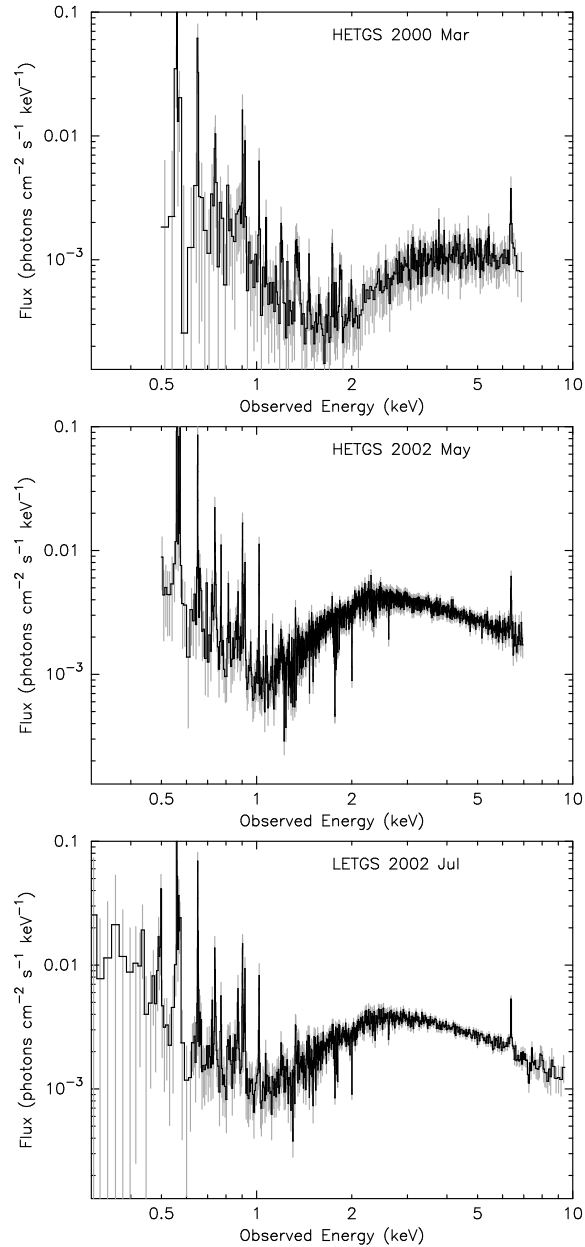

\begin{center}
\includegraphics[scale=0.30,angle=270]{f2a.ps}\\
\includegraphics[scale=0.30,angle=270]{f2b.ps}\\
\includegraphics[scale=0.30,angle=270]{f2c.ps}
\end{center}
\caption{Time-averaged, broad-band spectra for each of the {\it Chandra}
observations of NGC~4151 discussed here 
(MEG data only for the HETGS observations). In each case, the data 
have been corrected for the appropriate instrumental response, and 
(for clarity) rebinned such that each bin contains at least 30 photons. 
(Note this process clearly smooths out a number of the sharp 
emission and absorption features discussed later in the text). Error bars are shown in gray.
The low energies the spectra are dominated by emission lines, and are almost 
identical at all epochs. However above 1~keV, the emission is dominated by 
the nuclear continuum. This is very similar in both form and intensity for 
the two observations in 2002, but stronger and of a different shape to that 
observed in 2000.\label{fig:broad-spec}}
\end{figure}

\begin{figure}
\begin{center}
\includegraphics[scale=0.50,angle=270]{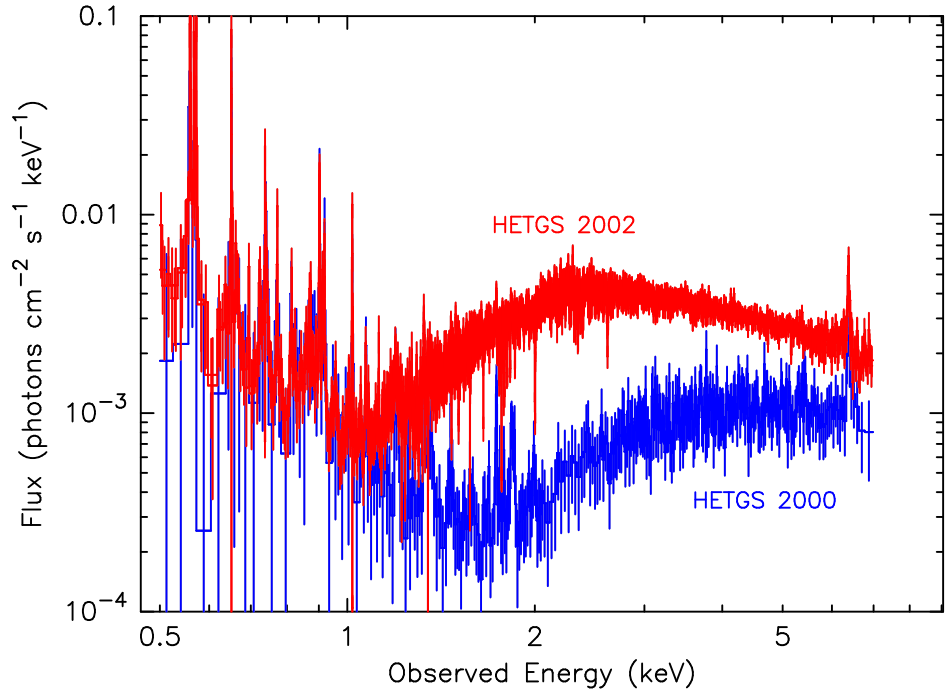}
\end{center}
\caption{A comparison of the 2000 (blue) and 2002 (red) HETGS data of NGC 4151.
As in Fig.2, the data have been rebinned such that each bin contains at least 
30 photons.
During both epochs, the flux at energies $<$ 1 keV is dominated by unocculted
line emission, which does not appear to have varied in flux.
\label{fig:broad-spec_v2}}
\end{figure}

Historically, the intrinsic X-ray continuum in NGC 4151 has been found to be heavily absorbed
at energies $\lesssim$ 1 keV  
(e.g., George et al. 1998, and references therein). 
To illustrate this,
in Fig.~\ref{fig:broad-spec} we show the time-averaged, 
broad-band spectra obtained during the three epochs.
For clarity of display only the MEG data are shown for the HETGS 
observations, and in all cases the data have been rebinned 
such that each bin contains at least 30 photons. Each data set 
has been corrected for the appropriate instrumental response. As a result of the
soft X-ray absorption,
absorption lines from abundant elements such as C, N and O cannot be detected. 
In fact, the bound-free
edges from O~VII, O~VIII, Ne~IX, and Ne~X, which were typically the signature of warm absorbers in the {\it ASCA} era
(e.g. Reynolds 1997), are so deep that they are blended together and produce  
a smooth curvature
at low resolution. In Fig~\ref{fig:broad-spec_v2} we show a direct comparison of the 2000 and 2002 HETGS spectra. In both
epochs, the soft band is dominated by emission lines which have not varied significantly in flux, while the hard band is both weaker and more 
heavily absorbed in the earlier spectrum. During both observations, the intrinsic continuum  
below $\sim$ 1 keV
is 
so heavily absorbed
that its contribution to the soft X-ray emission is negligible. However, with the 2002 spectra, we were able to detect absorption lines at energies
$>$ 1 keV from less abundant species, such as Mg, Si, and S. The ionic column densities measured from these lines,
in combination with the broad-band spectral curvature, allows us to constrain the ionization state and column densities
of the components of X-ray absorption.   

As noted in Section~\ref{Sec:Intro}, our analysis of the UV spectra will be presented separately. However the UV absorption 
provides crucial constraints for modeling the X-ray absorbers. To summarize the results of our UV analysis, we
found that the bulk of the UV absorption is associated with component D$+$E,
which has a velocity centroid of v$_{r}$ $=$ $-$ 491 ($\pm$ 8) km s$^{-1}$ (Weymann et al. 1997; Kraemer et al.  2001).
We were able to deconvolve D$+$E into three physically distinct components: a high covering factor component in which 
saturated OVI $\lambda\lambda$ 1031.9, 1037.6, N~V $\lambda\lambda$ 1238.8, 1242.8, and C~IV $\lambda\lambda$ 1548.2, 1550.8 lines
arise (which we will refer to as D$+$E{\it a}); a dense component responsible for the C~III$^{*}$ $\lambda$1175 absorption (D$+$E{\it b}); and a component with a lower covering factor
derived from the P~V $\lambda\lambda$ 1117.8, 1128.1 lines (D$+$E{\it c}). 
We also detected S~IV $\lambda$1062 absorption at v$_{r}$ $\approx$ $-$215 km s$^{-1}$, which suggests that the gas
associated with UV component E$^{'}$ is more highly ionized than that modeled by Kraemer et al. (2001). Each of these 
components contributes to the X-ray absorption.
The UV constraints for the X-ray modeling are listed in Table 1.

\begin{figure}
\begin{center}
\includegraphics[scale=0.50,angle=270]{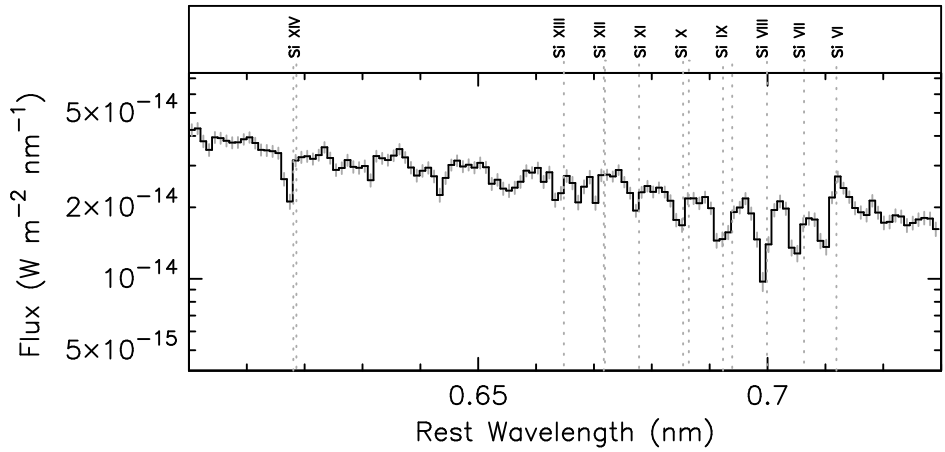}
\end{center}
\caption{A section of the combined MEG/HEG spectrum showing the lines
from Si~VI to Si~XIV using a binsize of 0.01\AA. The Si~XIV appears deeper than Si~XIII and the lower ionization states
appear to peak near Si~VIII, which strongly suggests that there are at least two absorbers with
differing ionization states. Also, note that the lines appear systematically blue-shifted,
similar to the UV lines, which suggests mass outflow.
 \label{fig:si_array_v1}}
\end{figure}

\begin{figure}
\begin{center}
\includegraphics[scale=0.50,angle=270]{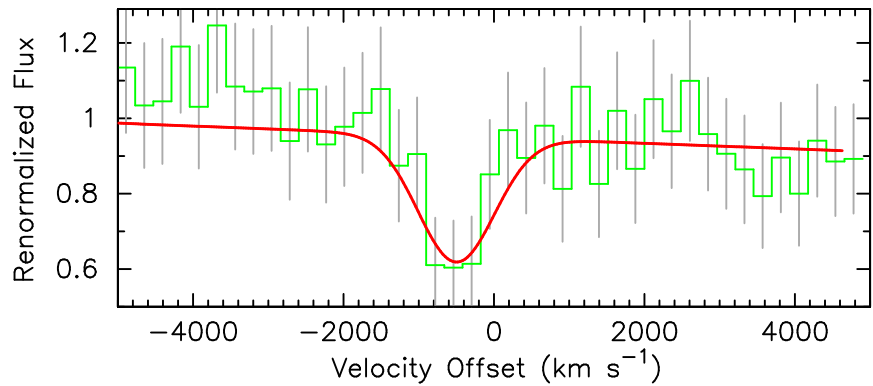}
\end{center}
\caption{The Si~XIV line profile from the 2002 HETGS spectrum is shown as 
a function of velocity. 
The binsize is 0.005 \AA, which is 1$\sigma$ for the HEG resolution, however both 
MEG and HEG are combined, so model is smoothed by MEG resolution. The uncertainties are
shown via the grey lines. 
We have overlaid a
Gaussian profile as a guide to demonstrate the velocity offset (v$_{r}$) with respect to
systemic. We determine v$_{r}$ to be $\sim$ $-$500 km s$^{-1}$.
 \label{fig:si14_vel}}
\end{figure}

From the S, Si, and Mg absorption lines detected
in the HETG spectrum, there is strong evidence for highly ionized gas, in addition to the lower ionization gas
that may be associated with the UV absorbers. This is clearly
seen in the Si array (Fig.~\ref{fig:si_array_v1}), which shows strong Si~XIV $\lambda$ 6.1804, 6.1858  and inner shell lines such as 
Si~X $\lambda\lambda$ 6.854, 6.864 and those from lower states of ionization, but has relatively weak 
S~XII $\lambda\lambda$ 6.717, 6.719 and Si~XI $\lambda$ 6.778. Although the Si~XIII $\lambda$ 6.648 line is affected by
unocculted emission from the He-like triplet, its weakness relative to Si~XIV is further evidence for absorbers of 
different ionization states, rather than a single, increasingly neutral column of gas. Therefore, we included 
a very highly ionized
absorber to account for the H-like lines (X-High).  Since the inner shell lines are from ionization states that can co-exist with the 
Li-like ions of C, N, and O, we assume that they arise in component D$+$E{\it a}.
While there are no UV constraints on the covering factor ($C_{f}$) for X-High, the depths of
the H-like lines suggest that $C_{f}$ is fairly large, otherwise these lines would be significantly diluted  
(see Turner et al. 2005). Hence we assumed $C_{f}$ $=$ 1.0 for this component. While we were only able to constrain 
the radial velocities for the S~XVI, Si~XIV and Si~XIII lines, we found v$_{r}$ for the high ionization gas to be
consistent with that of D$+$E (see Fig~\ref{fig:si14_vel}). The H-like lines appear to be narrow, although 
accurate measurements of their velocity dispersions are impossible due to
partial filling by neighboring emission lines. Nevertheless, we assumed a dispersion 
$\sigma_{v}$ $=$ 50 km s$^{-1}$ or roughly the thermal width for these ions in gas at a temperature of
$\sim$ few $\times 10^{6}$K, with the caveat that the intrinsic widths may indeed be substantially greater.

\subsection{The Photoionization Models}
\label{Sec:phot}

The photoionization models for this study were generated using the Beta 5
version of Cloudy (G. Ferland 2003, private communication), which includes
estimated $\Delta$n$=$0 dielectronic recombination (DR) rates for the M-shell states of Fe and
the L-shell states of the third row elements (Kraemer, Ferland, \& Gabel 2004). The
absorbers were modeled as single-zoned slabs of constant-density atomic gas, irradiated by the 
central source. We assumed
roughly solar elemental abundances (e.g. Grevesse \& Anders 1989) and 
the absorbing gas was assumed to be free of cosmic dust. The logs of the elemental abundances, relative to H 
by number, are
as follows: He: $-1.00$, C: $-3.47$, N: $-3.92$, O: $-3.17$, Ne: $-3.96$,
Na; $-5.69$, Mg: $-4.48$,  Al: $-5.53$, 
Si: $-4.51$,  P: $-6.43$, S: $-4.82$,  Ar: $-5.40$, 
Ca: $-5.64$,  Fe: $-4.40$, and Ni: $-5.75$.
As per convention, the models are parameterized in
terms of the ionization parameter $U$ (see Section~\ref{Sec:Intro}), and total hydrogen column density, $N_{H}$. 
 
From our STIS spectrum, we measured an extinction-corrected UV flux at 1450\AA~of 
$\sim 7.7\times10^{-26}$
ergs cm$^{-2}$ s$^{-1}$ Hz$^{-1}$ for an extinction $E_{B-V} = 0.04$ (Kriss et al. 1995). 
The 2--10 keV flux in the 
{\it Chandra} spectra was 
$\sim 1.9\times10^{-10}$ 
ergs cm$^{2}$ s$^{-1}$. Due to the heavy absorption, it is not possible to
tightly constrain the intrinsic X-ray photon index. Although Zdziarski, Johnson \& Magdziarz (1996), using combined
{\it ROSAT}, {\it Ginga}, and {\it Gamma Ray Observatory}/OSSE data, derived a $\Gamma = 1.8$, Yaqoob \& Warwick (1991) found that, like most Seyferts,
NGC 4151 shows a anti-correlation between spectral slope and intensity. Therefore,  
we assumed 
$\Gamma$ $=$ 1.5, which is consistent with the 2002 data and previous low-flux state values. Based on this index, the unabsorbed flux is 
$\sim 2.7\times10^{-28}$
 ergs cm$^{-2}$ s$^{-1}$ Hz$^{-1}$ at 0.5 keV.
Using these values, we modeled the spectral energy distribution (SED) as a broken power law of the form $L_{\nu} \propto 
\nu^{\alpha}$ as follows: $\alpha = -1.0$ for energies $<$ 13.6 eV,
$\alpha = -1.3$ over the range 13.6 eV $\leq$ h$\nu$
$<$ 0.5 keV, and $\alpha = -0.5$ above 0.5  
keV. We included a low energy cut-off at $1.24 \times 10^{-3}$ eV (1 mm) and a high energy cutoff
at 100 keV. The model SED is similar to that used in Kraemer et al. (2001).
The luminosity in ionizing photons is  
$Q = 1.1\times10^{53}$ photons s$^{-1}$. 
Note that, while the soft X-ray break could occur at lower energy,
we found that a higher energy break causes an overprediction of the soft-X-ray continuum.    

As described in the previous section, there is evidence for five distinct components with
non-negligible X-ray opacities. Given the large covering 
factors of components D$+$E{\it a}, D$+$E{\it b} and E$^{'}$, the most straightforward geometry places
the absorbers over a range in radial distances, with the more distant ones illuminated by
an ionizing continuum that has been filtered by the intervening absorbers (a similar model was used for NGC 3516;
Kraemer et al. [2002]). The only way to constrain the radial distance of a component is by determining both
$n_{H}$ and $U$, which we were only able to do for components D$+$E{\it b} and E$^{'}$ (as we will show in Paper II). 
However, it is plausible that the X-High and D$+E${\it a} lie closest to the 
central source, since they are the most highly ionized. Also, in our previous study of NGC 4151 (Kraemer et al. 2001), 
we only found evidence for a small column ($\sim$ 10$^{20}$ cm$^{-2}$) of high
ionization gas in the shadow of component D$+$E. 
Therefore, we assumed that the
absorbers were ordered in increasing radial distance as follows: X-High, D$+$E{\it a}, D$+$E{\it b},
and E$^{'}$. Since D$+$E{\it c} has a lower $C_{f}$, we assumed that it was co-located with
D$+$E{\it b}, i.e., only screened by the two higher ionization components, but did not include its affect on the
filtered continuum used to model E$^{'}$. 

We generated a grid of photoionization models  
for a range of $U$ and $N_{H}$ values, beginning with X-High and following with associated grids for the screened components, using the
transmitted continuum of each intervening component as the incident continuum for the next in radial order. The tables of ionic column 
densities predicted by Cloudy were used to construct
a grid of high-resolution spectra, including the effects of bound-free 
absorption (Verner et al. 1996), and resonance and inner shell 
absorption lines (e.g., Behar \& Netzer 2002).  The spectral grids were used as an 
input to {\tt xspec} in order to fit the broad-band X-ray spectrum.
Our initial set of models, based on the constraints listed in Table 1, slightly underpredicted the 
X-ray absorption at energies $\lesssim$ 0.7 keV. Since the  
soft X-ray emission in these spectra are dominated by 
the emission-line gas, only a small fraction of the emission at these energies can be from the transmitted
continuum (see Fig~\ref{fig:broad-spec_v2}).
Since the UV and X-ray continuum regions in NGC 4151 are suspected to have different sizes (Crenshaw et al. 1996; Edelson et al. 1996), using the
UV lines may lead us to underestimate C$_{f}$ for the X-ray source.
Therefore, we allowed the covering
factor (Table 1) to be $\sim$ 20\% higher, which improved the fit without the overprediction of the UV ionic column densities that would have
occurred if the higher soft-X-ray opacity were the result of additional low-ionization gas. 

The presence of a strong, narrow Fe K$\alpha$ (6.4~keV) and, perhaps, several other emission lines in the spectrum 
(e.g., see Figs. 2, 3) is indicative of fluorescence within  
optically-thick material within the circumnuclear regions following illumination by the 
primary X-ray continuum. Thus we have also included such a `reflection component'
(e.g. George \& Fabian 1991; Matt et al. 1991)
with our modeling.
Unfortunately the location, geometry and ionization-state of such circumnuclear material 
is not very well constrained by current data (for any object). 
Thus neither are the precise details of the reflected spectrum 
(which included both fluorescent lines and a scattered continuum). 
However, besides offering an explanation for the  Fe K$\alpha$ emission, the only significant 
effect of such a component for the results reported here
is a slight flattening of the observed continuum at high energies
($\gtrsim3$~keV). 
Thus we have simply assumed a `standard' reflection component
where the intensity of the reflected continuum is linked to the observed strength of 
the Fe K$\alpha$ line (using the calculations of George \& Fabian 1991).
Along with an incident primary continuum of $\Gamma = 1.5$, we assumed that the 
material subtends $2\pi$~sterad as seen from the central source, that the material 
has cosmic abundances, and is far from the central source and hence is effectively neutral. 
Our model also includes fluorescent lines from other abundant elements from the calculations
of Matt, Fabian \& Reynolds (1997). 

\subsection{Model Results}
\label{Sec:Results}

\begin{figure}
\begin{center}
\includegraphics[scale=0.50,angle=270]{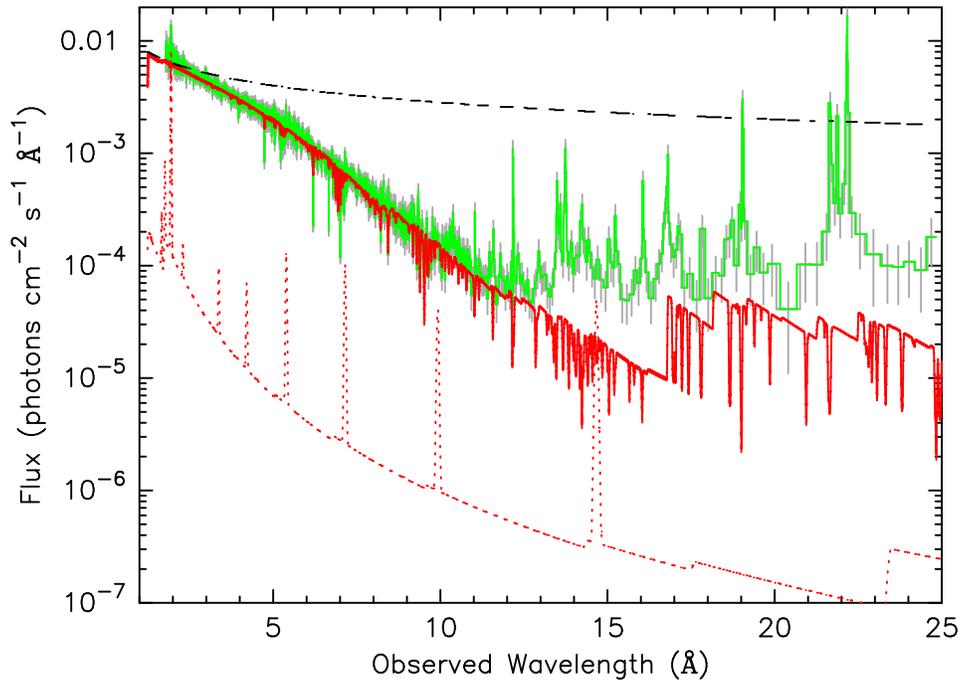}
\end{center}
\caption{The model absorbed continuum (red), including the reflection component (dotted
red line),
is shown versus the 2002 HETGS data (green). At wavelengths shorter than $\sim$
12 \AA~($\sim$ 1 keV), 
the model reproduces the curvature
in the observed continuum, while at longer wavelengths the absorber/transmitted continuum
lies below the observed points, indicating that the soft flux is dominated by
unocculted emission.
\label{fig:broad-spec_model02}}
\end{figure}

\begin{figure}
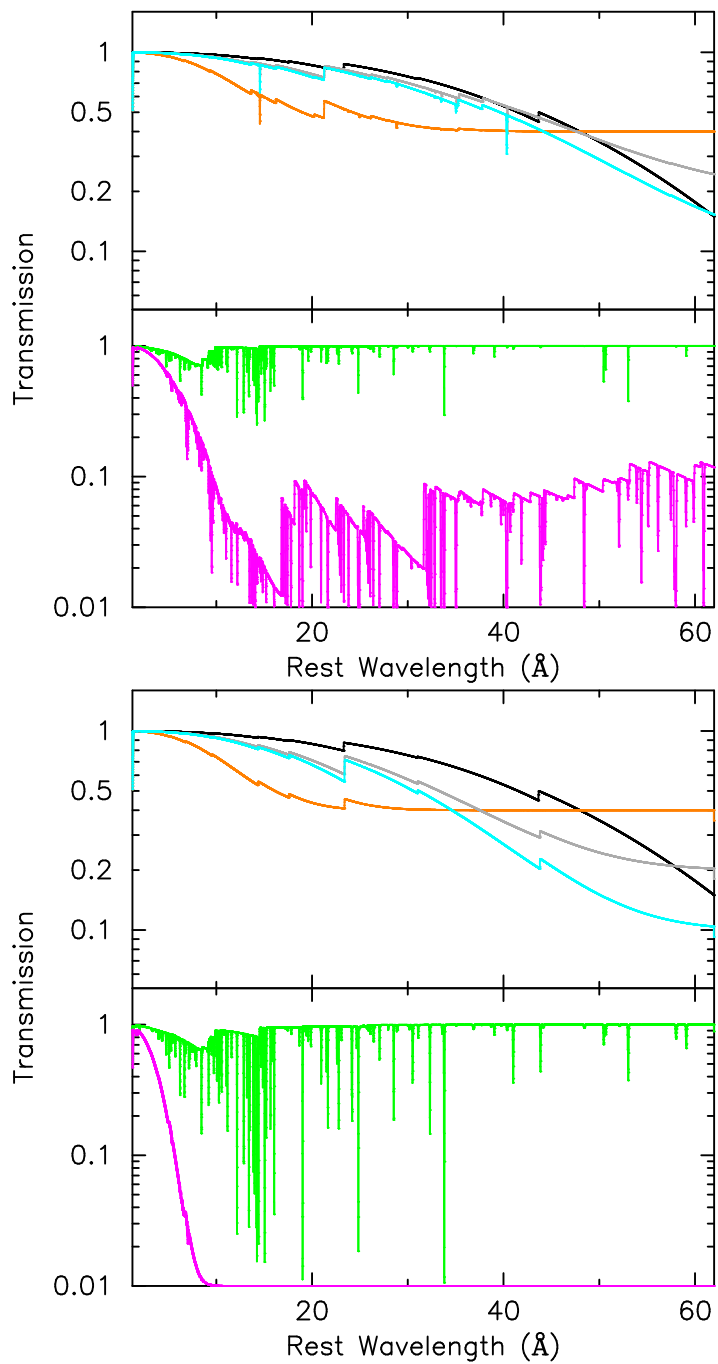

\begin{center}
\includegraphics[scale=0.50,angle=270]{f7a.ps}\\
\includegraphics[scale=0.50,angle=270]{f7b.ps}
\end{center}
\caption{The upper panel shows the 
broad-band transmission curves for each of the model components 
for the 2002 epoch discussed in the text. 
The lower panel shows the predicted curves for the 2000 intensity state. 
In all cases E$^{'}$ is grey, D$+$E{\it c} is orange, D$+$E{\it b} is blue, 
D$+$E{\it a} is magenta, and X-High is green. The Galactic component is shown in black. Note 
that components  E$^{'}$, D$+$E{\it c} and D$+$E{\it b} only 
partially cover the source, resulting in their curves flattening towards longer 
wavelengths due to the unocculted fraction.
\label{fig:opacity}}
\end{figure}

\begin{figure}
\begin{center}
\includegraphics[scale=0.50,angle=270]{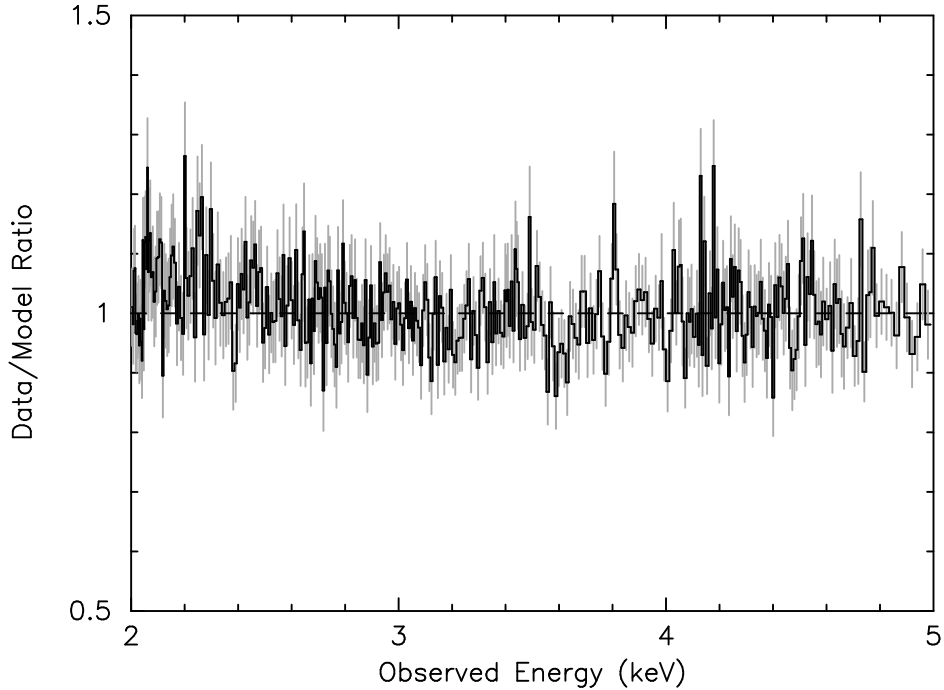}
\end{center}
\caption{
The $\pm1^{\rm st}$-order MEG (only) data divided by the 
model for the 2002 obs. The data are plotted such that each bin contained 
at least 200 photons for clarity.
A few unmodeled features are present, and there is an indication of a 
small systematic excess of the data compared to the data towards the 
lowest energies. However the effect is at a level of $\lesssim$10\%, and 
hence at a similar level and sense as the current uncertainties in the 
calibration associated with the Ir coating on Chandra's mirrors.
Furthermore, NGC 4151 does exhibit some variability during the 
observation, yet here we restrict our analysis to the time-averaged 
spectrum. With these facts in mind, we consider our model to provide 
an adequate match to the overall spectral curvature in the 
time-averaged spectrum.
\label{fig:2_5_data_model}}
\end{figure}

With the intrinsic continuum and reflection components held fixed,
we generated a set of absorber models to fit the broad-band spectral properties. We deemed the modeling to be
successful if the combined effects of the absorption and reflection provide a satisfactory fit
to the spectrum at energies where the emission is dominated by the transmitted continuum, i.e. $\gtrsim$ 1.5 keV,
and the UV-derived constraints are met. The parameters of our best fitting models are listed in  
Table 2 and the predictions of ionic column
densities are listed in Table 3. In Table 1, we compare the predicted ionic column densities to the UV constraints.
The model is shown versus
the combined MEG and HEG spectra in Fig.~\ref{fig:broad-spec_model02}, while the opacities of the individual components are shown in
Fig.~\ref{fig:opacity}. The curvature seen above $\sim$ 1 keV is primarily due to the bound-free edges from O~VI (inner shell), O~VII, and O~VIII 
edges predicted by the component D$+$E{\it a} and the Ne~X and Fe~XVII--XX edges from X-High. 
Note that a different set of solutions would be obtained if the radial locations of X-High and D$+$E{\it a} were switched, due to the effect of
the deep soft-X-ray absorption from D$+$E{\it a}. Although we cannot rule out this scenario, we found that reversing the order of the models provides a
less satisfactory fit to the curvature below $\sim$ 2 keV, due to the overprediction of the O~VIII column density in the high ionization
component which results when the H- and He-like Mg, Si, and S columns are held fixed. Fig~\ref{fig:2_5_data_model} shows the data/model ratio 
in the  2--5 keV band, where we have the best constraints on the transmitted continuum.
After grouping the data such that there are at least 50 photons in each 
bin, a simultaneous comparison between the MEG and HEG data and our model 
in the 2--5 keV band gives a $\chi^2$-statistic of 1694 for 1708 degrees of 
freedom (i.e. a reduced $\chi^2$ value of 0.992).
Component D$+$E{\it a} also produces the 
bulk of the absorption below 1 keV, although there is
some contribution from the lower ionization intrinsic absorption, mostly due to He~II bound-free absorption, and 
the Galactic neutral column ($N_{H}$ $=$ 2.0 x 10$^{20}$ cm$^{-2}$; Murphy et al. 1996). Note that the model is not tightly constrained by the fit at 
energies $<$ 1 keV, since the soft-X-ray emission is dominated by the extended emission-line gas, except in the sense
that the transmitted continuum must be weaker than the observed X-ray emission over the entire soft band.

\begin{figure}
\begin{center}
\includegraphics[scale=0.60,angle=270]{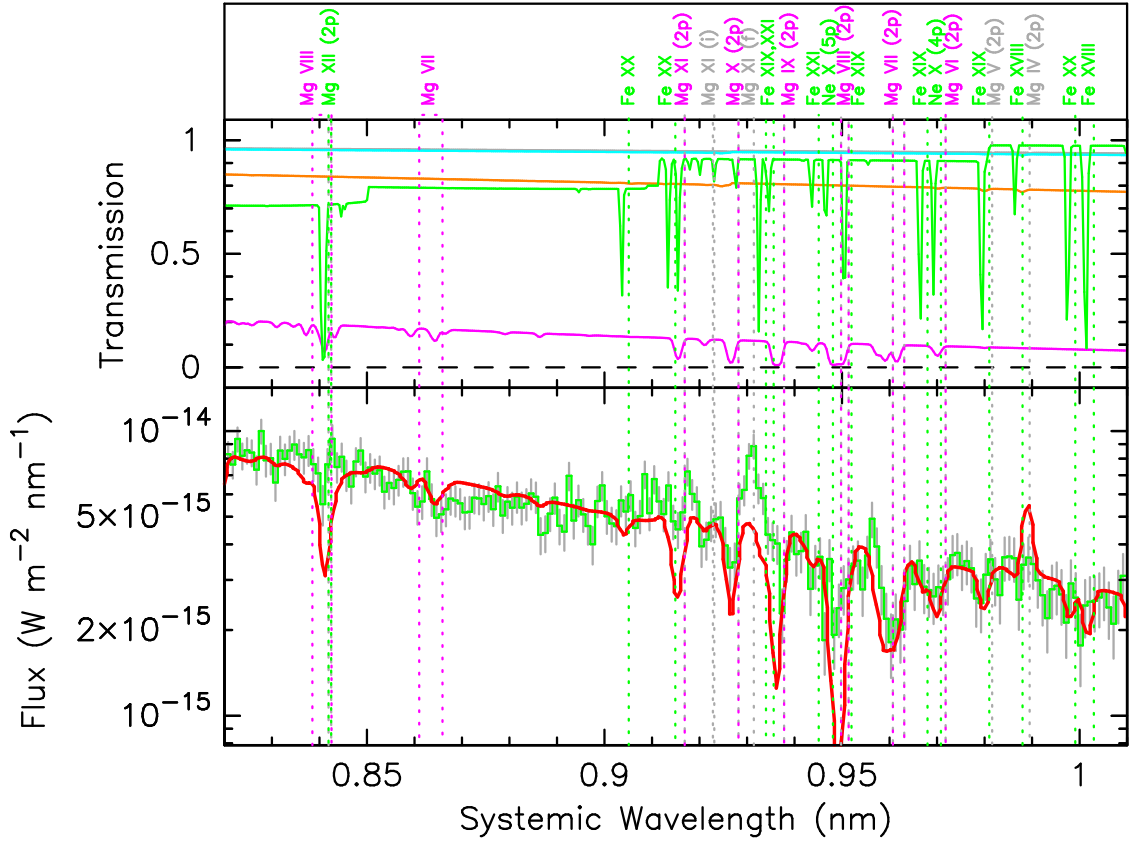}
\end{center}
\caption{The region of the time-averaged spectrum obtained 
from the 2002 HETGS observation of NGC~4151 showing the Mg lines.
The lower panels shows (in green) the coadded MEG and HEG  
$1^{\rm st}$-order spectra (after instrumental corrections 
have been applied, and assuming a systemic $z=0.003319$) 
using 0.01 \AA~bins. We note that absorption lines due to transitions to the $3p$ level are 
also evident in the data for Mg~VII.
The model spectrum discussed in the text is shown by the 
red curve (after smoothing by the instrumental resolution).
The upper panels show the intrinsic transmission curves of the individual 
components to the model. As in Fig.~\ref{fig:opacity},
D$+$E{\it c} is orange, D$+$E{\it b} is blue, 
D$+$E{\it a} is magenta, and X-High is green.
Also shown are the locations of several absorption 
lines (color-coded for the component dominating the feature).
\label{fig:arraymg}}
\end{figure}

\begin{figure}
\begin{center}
\includegraphics[scale=0.60,angle=270]{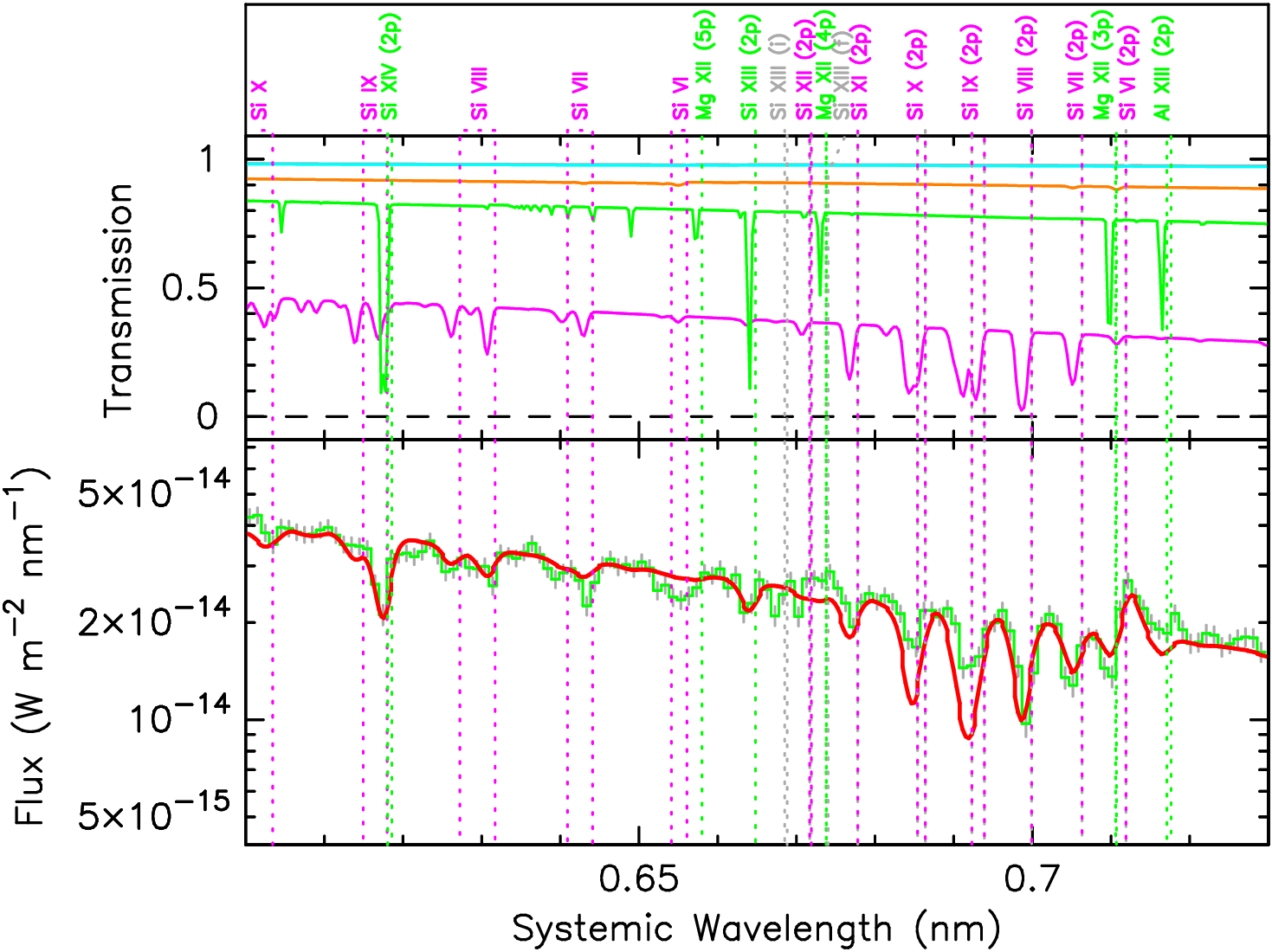}
\end{center}
\caption{The regions of the time-averaged spectrum obtained 
from the 2002 HETGS observation of NGC~4151 showing the Si lines.
Color-code and model parameters are the same as described for
Fig~\ref{fig:arraymg}.
Absorption lines due to transitions to the $3p$ level are 
also evident in the data for Si~VII and Si~VIII.
\label{fig:arraysi}}
\end{figure}

\begin{figure}
\begin{center}
\includegraphics[scale=0.60,angle=270]{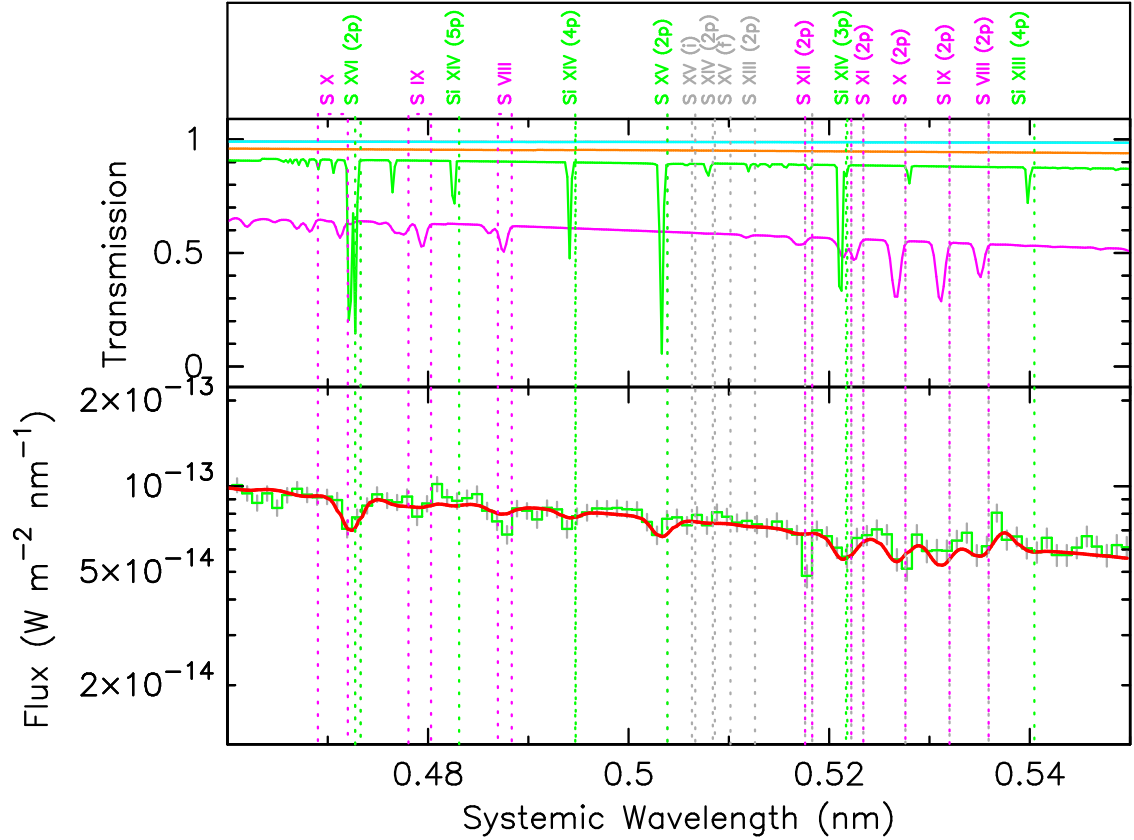}
\end{center}
\caption{The regions of the time-averaged spectrum obtained 
from the 2002 HETGS observation of NGC~4151 showing the S lines.
Color-code and model parameters are the same as described for
Fig~\ref{fig:arraymg}.
Absorption lines due to transitions to the $3p$ level are 
also evident in the data for S~VIII and S~IX.
\label{fig:arrays}}
\end{figure}

Next, we examined the predicted X-ray absorption line spectrum. Since the
ionic columns densities ($N_{ion}$) were determined by matching the broad-band spectral features, the absorption-line fit 
provides a consistency check for the models. We generated the line
profiles based upon the $N_{ion}$'s, using the $\sigma_{v}$'s assumed for each component. As noted above, since the strong
resonance lines of H- and He-like C, N, O, and Ne have energies $<$ 1.2 keV, their imprint against the
transmitted continuum is swamped by the extended soft emission, and our study is limited to
the H-, He-like and inner shell lines of Mg, Si, and S. 
In Figs.~\ref{fig:arraymg},~\ref{fig:arraysi}, and ~\ref{fig:arrays}, we show portions of the time-averaged 
spectrum obtained by the HETGS in 2002. These regions were selected 
so as to contain all the $1s\rightarrow2p$ transitions for H- through 
F-like ions of Mg, Si and S.
The combined MEG and HEG $1^{\rm st}$-order 
data are shown (after correction for the 
appropriate instrumental responses) and are displayed using 
0.01\AA~bins (approximately the ``1$\sigma$'' width of the spectral 
resolution of the MEG). The model predictions 
are overlaid on the data (the emission features are from the reflection component model and the spectra are consistent with
K-shell features from the disk). The Mg lines, with energies $<$ 1.5 keV, are in a region of the X-ray spectrum where the
transmitted/absorbed continuum begins to become weak relative to the extended emission (see Fig~\ref{fig:broad-spec_model02}). Therefore, it is difficult to determine the
actual depths of most of the absorption lines
and, our model appears to overpredict several of them. However, the fit is good for Mg~VI (combined with Fe~XIX and Ne~X) 
and Mg~VII, and the features from the other Mg ions appear to be present, although shallower than the model predictions. For the S and Si arrays, 
where the transmitted/absorbed continuum dominates, the model provides a better overall 
fit to the absorption lines. In both cases,
the model, primarily X-High, predicts the strong H-like and weaker He-like lines observed. The predictions for the inner shell lines
are best assessed with the Si array (Fig.~\ref{fig:arraysi}).  
Generally, the 
model (D$+$E{\it a}) provides a fairly good fit, in the sense that all of the predicted lines are present in the 
spectrum. Nevertheless, Si~IX and Si~X are 
somewhat overpredicted and Si~VI is underpredicted. We tested whether increasing $N_{H}$ or decreasing $U$ for D$+$E{\it a} would
drive Si to sufficiently lower ionization, but in doing so we overpredicted $N_{PV}$ from this component. We also tested the 
effect varying $\sigma_{v}$. While the overpredictions are mitigated by a assuming $\sigma_{v} = 85$ km s$^{-1}$, the fit becomes worse
for the underpredicted lines, and the opposite occurs for $\sigma_{v} = 285$ km s$^{-1}$. 
However, the 
discrepancy corresponds to over/under-predictions of the ionic column densities by factor of $<$ 2. These could easily
be the result of inaccuracies in the DR rates used in Cloudy, since the analytic form used for the third row elements does
not include structure within the L-shell (see Gu 2003).         
 
\subsection{Comparison with 2000 HETGS Spectrum}
\label{Sec:HETG2000}

One of the central questions in understanding the nature of intrinsic absorption and mass outflow in AGN is
whether temporal variations in the spectral signature of the absorbers, including the appearance/disappearance
of individual components, result from changes in ionization state or bulk motion (see Crenshaw et al. 2003).
The 2002 LETGS spectrum found the X-ray continuum to be roughly the same flux and possess similar spectral characteristics as the
2002 HETGS observations (Fig~\ref{fig:broad-spec}). This suggests that the physical conditions in the X-ray absorber
did not change between these observations; indeed, our model provides a reasonable fit to the LETGS data. 
The 2000 HETGS observations found NGC 4151 in a lower flux state than in 2002 (see Fig~\ref{fig:broad-spec_v2}), with 
a steeper curvature above $\sim$ 1 keV, consistent with an increase in the opacity of the absorber, as expected 
if the gas had become more neutral in response to the drop in ionizing radiation. Noting that some of the drop in the X-ray
flux must be due to the increased absorption, we estimate that the intrinsic flux was a factor of $\sim$ 2 lower
during the 2000 observation. Starting with X-High, we regenerated our set of models for this lower flux state, using the
same SED and holding
$N_{H}$ fixed, and 
compared the results to the 2000 HETG spectrum. The model was unsuccessful in matching both the strength of the absorption
and the curvature in the 2--5 keV range. We found little improvement when we decreased $U$ for X-High by a factor of 3,
which exceeds the maximum allowed by the data. 

\begin{figure}
\begin{center}
\includegraphics[scale=0.50,angle=270]{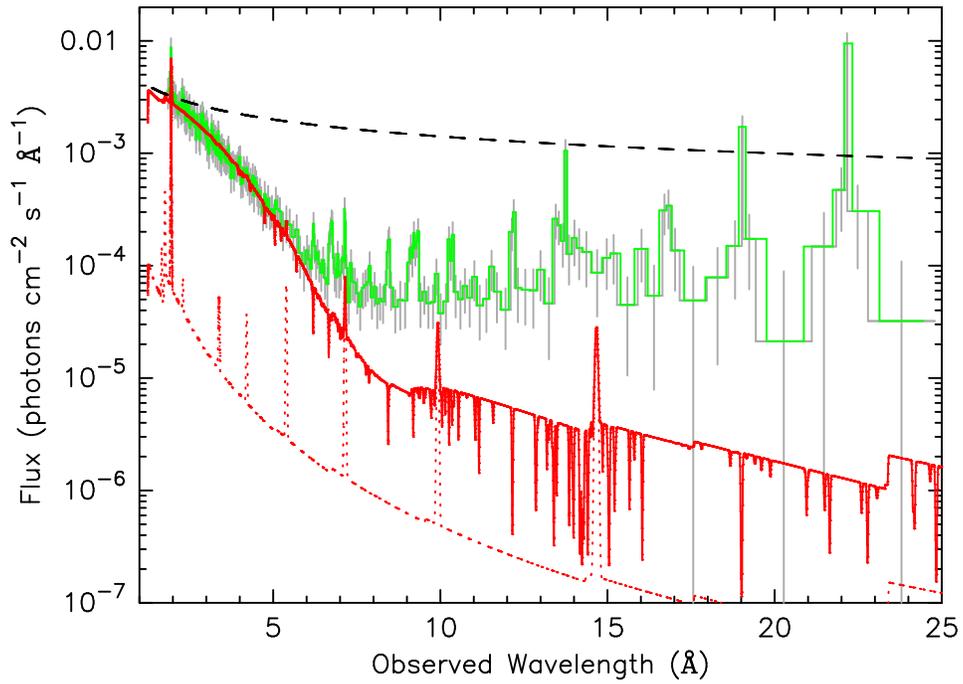}
\end{center}
\caption{The model absorbed continuum (red), including the reflection component (dotted red line),
is shown versus the 2000 HETGS data. For the model the ionizing flux was decreased by a factor
of 2 and the column density of component
D$+$E{\it a} was increased by a factor of 3 compared to the 2002 model.  
\label{fig:broad-spec_model00}}
\end{figure}

The mismatch between the models and the 2000 spectrum indicates that difference in the X-ray absorption cannot result 
simply from changes in ionization state. Instead, the total column density of the absorbers must have been greater in
2000, as suggested by Schurch \& Warwick (2002) based on their analysis of the 2000 May {\it ASCA} observations. To test this, we let column density 
for D$+$E{\it a} be a free parameter and obtained a rough fit to the 2--5 keV curvature
for log$N_{H}$ $=$ 22.93. The opacities of the low-state versions of X-High and D$+$E{\it a} are shown in Fig~\ref{fig:opacity}.  
Under these conditions, the other screened components become effectively neutral, therefore we replaced them
with a neutral column of log$N_{H}$ $=$ 21.7. The model is shown against the data in Fig~\ref{fig:broad-spec_model00}, and it 
provides a reasonable fit to the broad-band spectrum, albeit with the
assumption that the transmitted continuum has a negligible contribution to the observed flux below $\lesssim$ 2 keV.
We suggest that this is evidence for a combination of the response to the drop in ionizing flux and bulk motion across 
our line of sight. We will pursue this in Paper II via a detailed comparison of the 1999 and 2002 STIS spectra.

\subsection{Physical Conditions within the Absorbers}
\label{Sec:physics}

\begin{figure}
\begin{center}
\includegraphics[scale=0.60,angle=90]{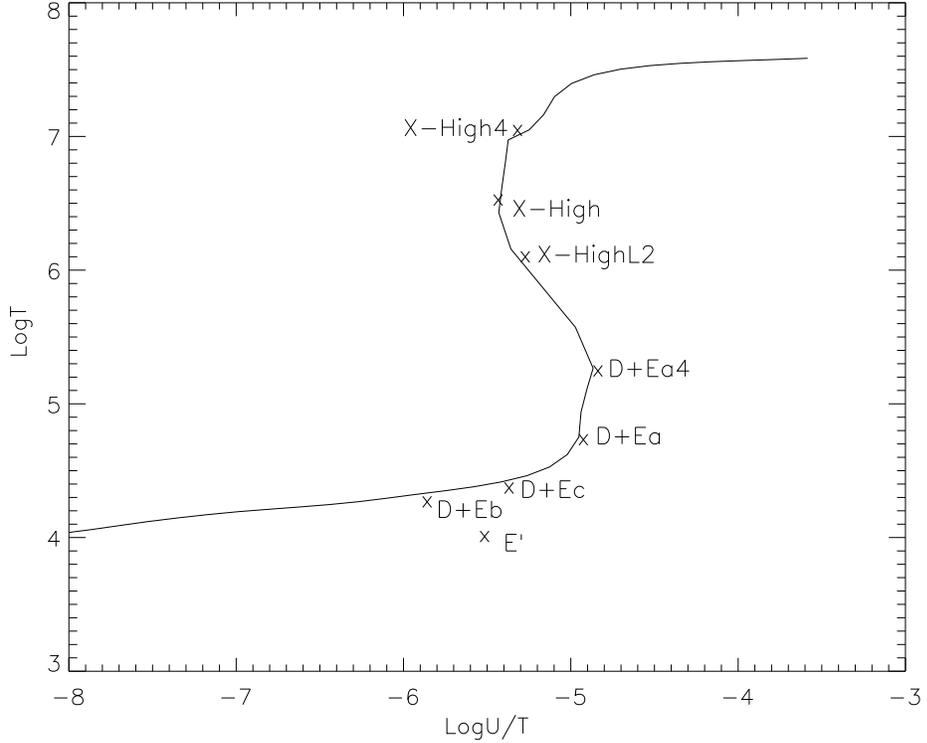}
\end{center}
\caption{Thermal Stability curve for optically-thin models generated with the
same SED used for the absorbers. The positively sloped sections, at low and high
temperatures, are the line-cooled and Compton -cooled regions, respectively.
As noted in the text, the vertical regions are quasi-stable, while the region with negative
slope is unstable to thermal perturbations. 
The values for each absorber component are shown (see text), as well as those for a source luminosity
4 times greater than during the 2002 observations (D$+$E{\it a}4 and X-High4) and 
2 times weaker, as during the 2000 HETGS observations (X-HighL2).
\label{fig:S-Curve}}
\end{figure}

Since the earliest UV (Boksenberg et al. 1978) and X-ray (Ives et al. 1976) observations, NGC 4151 has shown strong 
intrinsic absorption, which suggests that either 1) the absorbers are stable over relatively long periods
($\gtrsim$ 30 yrs) or that 2) the gas is constantly replenished. The thermal stability of photoionized gas is
illustrated via a thermal stability curve (``S-Curve''), which is a plot of the electron temperature, $T_{e}$, versus the ``pressure'' ionization parameter,
$U/T_{e}$. There are two regimes of ionization for which photoionized gas is stable to thermal
perturbations (see Krolik, McKee \& Tarter 1981; Krolik \& Kriss 2001): low ionization/temperature, when line cooling
is efficient, and at high ionization/temperature, when Compton processes dominate. The intermediate region
can be quasi-stable (when $T_{e}$ versus $U/T_{e}$ has a positive slope) or unstable (negative slopes) to thermal
perturbations.
Note that the form of the S-Curve is sensitive to the SED and the elemental abundances. Furthermore, if 
adiabatic cooling is included the maximum temperature can be considerably lower than the Compton
temperature (see Chelouche \& Netzer 2005).
In Fig~\ref{fig:S-Curve} we show the thermal stability curve derived for our model SED and abundances
and indicate the positions of the 5 component models. For the most part, these components are not clustered
near a single value of $U/T_{e}$, hence they are not in close pressure
equilibrium with one another, with the exception of D$+$E{\it c} and X-High, which may be coincidental
given the fact that we assumed a filtered continuum for the former. 

Not unexpectedly, the four UV components lie on the line-cooled
part of the curve, since the very fact that we detect absorption from Li-like CNO ions and lower ionization states 
requires there to be a non-negligible fraction of bound electrons. Note that these components, in particular E$^{'}$, lie
off the curve due to the fact that the models were generated with a filtered ionizing continuum (hence a different SED). The most highly ionized of these, D$+$E{\it a} lies 
near the transition between stable and quasi-stable sections of the curve. To examine its stability during high-flux
states, we generated a model for an ionizing flux 4 times that observed in 2002, similar to historic high-states
(see Kraemer et al. 2001). Although the high-state D$+$E{\it a} lies in the quasi-stable part of the curve, it would not
be thermally unstable. Therefore, if this component has remained in our line-of-sight to the
nucleus of NGC 4151, it would have always been seen in absorption, with a spectral signature that 
reflects changes in its ionization state in response to variations in the ionizing flux. 
 
In the current epoch, X-High resides in a quasi-stable region below the Compton-cooled regime. For the high-state model,
X-High is driven towards the Compton-cooled region, but remains in a quasi-stable state. We also examined the behavior
of this component for a flux state 2 times lower than the present epoch, similar to the 2000
HETGS observations. Notably, X-High becomes thermally unstable as the flux drops. This suggests a possible scenario
for the evolution of the absorbers. If the bulk of the absorbing gas originates in a highly ionized phase, similar to X-High, 
the gas could become thermally unstable during low-flux states. At these times, the gas may rapidly cool and recombine, until
its physical state resembles that of D$+$E{\it a}. If it has cooled/recombined sufficiently, it may remain thermally stable.
This is what might be expected if the flow originates as a hot thermal wind (Krolik \& Kriss 1995, 2001). The lower ionization UV
absorbers could also form as condensations out of the hot phase or might be ambient gas swept up by the flow. Alternatively, the low ionization
absorbers may have no physical connection to the high-ionization gas and simply lie along the same line-of-sight.

\section{Discussion}
\label{Sec:Discussion}

We have modeled the X-ray absorption in NGC 4151 with 5 independent components, at least three of which are observed at
roughly the same radial velocities. The absorbers span a wide range in ionization and are not in close pressure
equilibrium. Based on their large covering factors, the more distant absorbers must be screened by those closer to the 
central source. Furthermore, the evidence for bulk motion (Section~\ref{Sec:HETG2000}) suggests that there is a large non-radial
component to the outflow velocities, as might result if the source gas were rotating around the central black hole (e.g. 
the accretion disk).
Given the physical conditions predicted by our models, we can begin to probe the dynamics of
the outflow.

Based on the factor of $\sim$ 3 decrease in the column density of D$+$E{\it a}, we can estimate its transverse velocity.       
The size of the broad-line region in NGC 4151 is estimated to be $\approx 8.5 \times 10^{15}$ cm (K. Metzroth et al., in
preparation). Since the gas moved out of the line-of-sight over the 26 months between the two HETGS observations,
the transverse velocity $v_{T}$ $\gtrsim$ 1250 km s$^{-1}$. If $v_{T}$ is purely rotational, assuming a 
black hole mass of 10$^{7.1}$ M$_{\odot}$ (Peterson et al. 2004), D$+$E{\it a} must have originated at a radial
distance $R \lesssim 1.1 \times 10^{17}$ cm from the central potential. Note that this is on the same order as the
radial distance estimated from photoionization models of the 1999 STIS spectra (Kraemer et al. 2001), using a density 
determined from the strength of the C~III$^{*}$ lines, a point we address in more detail in Paper II.

\begin{figure}
\begin{center}
\includegraphics[scale=0.60,angle=90]{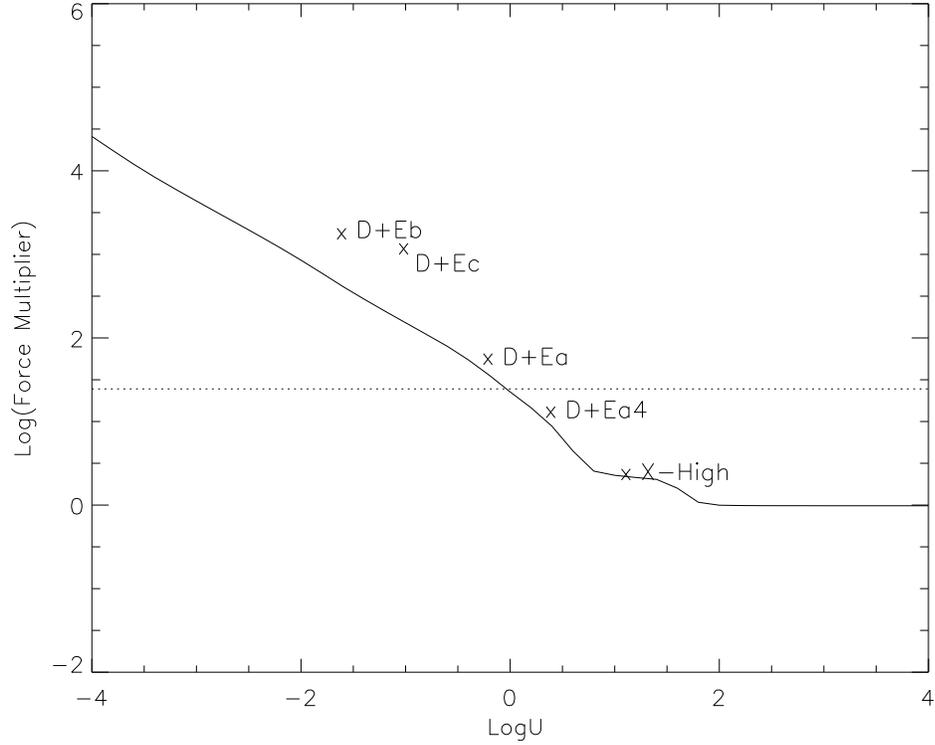}
\end{center}
\caption{The predicted Force Multipliers (FM) are shown as a function of $U$ for
optically thin models assuming the same SED used for the absorbers. The horizontal line shows the minimum value of FM
required to accelerate the gas  for 
a source radiating at 0.04L$_{E}$ (see text). 
The values for X-High and each component of D$+$E are shown, as well as the locations for 
for a source luminosity 4 times greater than during the 2002 observations (D$+$E{\it a}4) 
The screened models
lie off the curve since the fractional ionization as a function of $U$ is lower due to the
paucity of photons at the He~II Lyman limit..
\label{fig:FM_U}}
\end{figure}

Although the outflow is not purely radial, radiation pressure may play a role in accelerating the gas. The 
least efficient process by which
gas can be radiatively accelerated is the scattering of photons off electrons. Other sources of
opacity within the gas include line opacity (e.g, Castor et al. 1975),
bound-free transitions 
(Arav, Li, \& Begelman 1994; Chelouche \& Netzer 2001), and dust (e.g.
Konigl \& Kartje 1994); the latter is of minimal importance in the absorbing gas in NGC 4151, given the near absence of intrinsic 
extinction along our line-of-sight to
the nucleus (Crenshaw \& Kraemer 2005). The ratio of the line and bound-free acceleration to that of electron scattering is the ``Force
Multiplier'' ($FM$). For $FM =$ 1, the source must radiate at its Eddington Luminosity ($L_{E}$) to generate
an outflow. As such, $FM$ must be $\geq (L_{bol}/L_{E})^{-1}$, where $L_{bol}$ is the bolometric luminosity of the AGN, in order
to radiatively drive gas from the nucleus. For a typical SED, $L_{bol} \sim
9 \times \lambda~L_{\lambda}(5100 \AA)$, where $L_{\lambda}(5100 \AA)$ is the continuum luminosity at 5100 \AA~(Kaspi et al. 2000b).
For log$\lambda~L_{\lambda}(5100 \AA)$ $\approx$ 42.88 (Peterson et al. (2004), 
NGC 4151 is radiating at $\sim$ 0.04$L_{E}$, therefore radiative driving can only occur in gas with FM $\geq$ 25.
Fig~\ref{fig:FM_U} shows $FM$ as a a function of $U$ for models generated with our assumed SED. These values were computed for the
optically thin case, which is the case when radiative acceleration is most effective (the average $FM$ within a cloud decreases as the gas becomes optically thick;  
see Chelouche \& Netzer [2005]). Clearly, X-High is too highly 
ionized to be radiatively driven in NGC 4151. While the low-ionization components have
sufficiently high values of $FM$, D$+$E{\it a} is only marginally 
susceptible to radiative driving. Interestingly, while a substantial fraction, perhaps most, of the gas cannot be 
radiatively driven, the absorbers seem to possess similar outflow velocities. 

Another mechanism for the acceleration of gas  
is via 
thermal winds
(e.g., Begelman, McKee, \& Shields, 1983), 
arising either from an accretion disk or
the putative molecular torus (Krolik \& Begelman 1986, 1988) and subsequently accelerated 
along thermal pressure gradients (Balsara \& Krolik 1993; Chelouche \& Netzer 2005). One may
define an ``escape temperature'' for the gas, $T_{g} =
GMm_{H}/R_{0}k$, where $M$ is the central mass, $m_{H}$ is the mass of
a hydrogen atom, $R_{0}$ is the distance to the attracting mass, and
$k$ is Boltzmann's constant (see Crenshaw et al. 2003, and references
therein).  If $T > T_{g}$, a thermal wind will arise. Based on the predicted temperature,
one can determine the minimum radial distance at which such an outflow
can form, i.e., at a distance $R$ $\geq$ $R_{0}$. The most likely candidate for a thermal wind is the
component X-High. To determine the range of physically plausible radial distances,
we impose the constraint that the ratio of the
physical depth of the absorber, $\Delta R$ $=$ $N_{H}$/$n_{H}$, to its
radial distance, $R$, be less than unity.  
From Equation (1), $Q$
(see Section~\ref{Sec:phot}), and our predicted $U$ for X-High (Table
2), we find that 
$R \approx 1.6\times10^{20} n_{H}^{-1/2}$ cm.
For log$N_{H}$ $=$ 22.5 and $\Delta R/R$ $\leq$ 1, $R \leq 8.3\times10^{17}$ cm.
Cloudy predicts a temperature for X-High of $T_{e} = 3.0\times10^{6}$K. 
If we assume that any mass other than that of the central
black hole is negligible, 
$R_{0} \geq 7.1\times10^{18}$ cm. Therefore, X-High cannot have formed as a thermal wind under these
conditions. It is possible that the individual X-ray and UV components
are borne by a hotter wind, but it must be so highly ionized that it
produces no detectable absorption features. Such a wind would lie on
the Compton-cooled part of the thermal stability curve, where 
$T_{e} \sim 3\times10^{7}$K,
 and it would have formed at a radial distance 
$R \sim 6.7\times10^{17}$ cm. 
However, this places the wind at a larger distance
than that estimated for component D$+$E{\it a}, either by means of the
constraints from the bulk motion or from previous photoionization
models (Kraemer et al. 2001).

If the absorbers originate within the accretion disk, they may be accelerated via accretion-driven winds (Rees 1987). 
Models of disk-driven magneto-hydrodynamic (MHD) models predict that gas will flow outwards along (locally) open, rotating 
magnetic field lines (see Blandford \& Payne 1982; 
Emmering, Blandford, Shlosman, 1992).
Bottorff et al. (2000) applied MHD flows to the intrinsic absorption in Seyfert 1 galaxies.
Not only do such flows possess large non-radial velocities, but Bottorff et al. predicted that the most highly ionized
gas would exist at the smallest radial distances. As noted in Section~\ref{Sec:phot}, the relationship between ionization
state and relative radial distance is suggested by the large covering factors of the individual absorbers. 
For magnetically confined clouds in the presence of non-dissipative MHD waves, line widths can greatly exceed
the thermal widths, particularly if the magnetic field is in equipartition with the
gravitational energy density (Bottorff \& Ferland 2000, and references therein\footnote{Although Bottorff \& Ferland 
treat the case of broad emission lines, the same analysis applies to absorption lines.}). This scenario could explain 
the broad line profiles from component D$+$E{\it a} and provide a mechanism for pressure confinement if D$+$E{\it a} is co-located with X-High.
Therefore, although we cannot rule out a thermal wind origin and radiative acceleration may be important for the
lower ionization components, we suggest that the absorption in NGC 4151 is most consistent with an MHD flow.  
 
Independent of the mechanism driving the outflow, we can use the model parameters to constrain the mass outflow
rate. For a spherical outflow, the mass of the absorber is given by $M_{abs} = 4\pi r^{2} N_{H} m_{p} C_{g}$, where 
$m_{p}$ is the proton mass and $C_{g}$ is the global covering factor. $C_{g}$ can be constrained by comparing the
predicted emission-line fluxes to the observed line luminosities. The model for D$+$E{\it a} predicts a
flux $= 5.7 \times 10^{5}$ erg s$^{-1}$ cm$^{-2}$ for the O~VII
$\lambda$ 22.1 line (the forbidden line of the He-like triplet). The measured flux is 
$\approx 3.5 \times 10^{-13}$ erg s$^{-1}$ cm$^{-2}$, which yields a luminosity $\sim 7.3 \times 10^{39}$ ergs s$^{-1}$. 
For a radial distance of $r = 10^{17}$ cm, $C_{g} \sim 0.1$, which is an upper limit since much of the O~VII emission is 
extended (as we will discuss in Paper III). Thus, $M_{abs} \lesssim 0.3 M_{\odot}$. Note that this is the mass of D$+$E{\it a} alone; there could
be a similar amount of mass from X-High. The average mass loss rate is $\dot{M}_{abs} = M_{abs}/t_{c}$, where $t_{c} = r/v_{r}$ is the radial travel time.
If the flow originates at the origin ($r = 0$), we obtain $\dot{M}_{abs} \sim 0.005 M_{\odot}$ yr$^{-1}$ and a kinetic luminosity $1/2 \dot{M}_{abs} v_{r}^{2}
\sim 4 \times 10^{38}$ ergs s$^{-1}$. Assuming an efficiency of 0.1 (see Peterson 1997), $L_{bol}$ requires a mass
accretion rate of $\dot{M} \sim 0.001 M_{\odot}$ yr$^{-1}$ and, thus, based on this set of assumptions, the outflow and accretion rates
are comparable, similar to the case in NGC 3516 (Crenshaw et al. 2003). 
However, if the flow originates far from the continuum source, $\dot{M}_{abs}$ could be much greater, which would require the flow to be
continuously replenished (Netzer et al. 2003; Chelouche \& Netzer 2005).

\section{Summary}

We have used the {\it Chandra}/HETGS and photoionization models to probe the physical conditions in the
X-ray absorbing gas in the Seyfert 1 galaxy NGC 4151. Our models are further constrained by
covering factors, radial velocities, and ionic column densities derived from simultaneous UV spectra
obtained with {\it HST}/STIS and {\it FUSE}. We compared 
the 2002 HETG spectrum with one obtained during 2000 when NGC 4151 was in a lower flux state. The soft X-ray emission in both
epochs is dominated by emission lines, which have not varied in flux. Hence there is a negligible 
contribution from the transmitted continuum below 1 keV. 

We find evidence for at least two distinct X-ray absorbers: high-ionization gas (X-High),
in which absorption lines from H-like Mg, Si, and S arise, and a less ionized component, 
in which inner shell absorption lines from these elements form. We identify the latter with
the UV absorption at $-$ 491 km s$^{-1}$, component D$+$E in Kraemer et al. (2001). Interestingly,
X-High has a similar radial velocity. We have also
included three additional absorbers in our models: a high density component, D$+$E{\it b}, a lower covering
factor component, D$+$E{\it c}, both associated with the UV component at $-$491 km s$^{-1}$, and UV
component E$^{'}$. Since all the components other than D$+$E{\it c} cover $>$ 80\% of the continuum source, they must
screen each other from the ionizing radiation. Based on their relative opacities, we suggest that
X-High is closer to the central source than D$+$E{\it a}, followed by the additional UV components. Our model 
predicts both broad-band features and absorption lines in good agreement with the 2002 HETG spectrum.
The 
absorbers are not in pressure equilibrium with each other, which suggests (as does our screening scenario) that the individual
components are not co-located, although we cannot rule this out if magnetic confinement is important
(e.g. Gabel et al. 2005). While the models predict that each of the X-ray absorbers is thermally stable, X-High may become 
unstable if the continuum flux in NGC 4151 drops sufficiently. Hence, it is possible that lower ionization gas could
have formed from the high ionization component of the flow during previous low flux states. 

The parameters for X-High (log$U = 1.05$, log$N_{H} = 22.5$) are similar to those of the high ionization
absorbers detected in {\it Chandra} and {\it XMM-Newton} spectra of NGC 3783 (Netzer et al. 2003; Behar et al. 2003) and NGC 3516 (Turner et al.
2005), although the latter possesses an additional component with a much higher column density but small
covering factor. We find that X-High is too highly ionized to be radiatively accelerated in a source as 
sub-Eddington as NGC 4151. The presence of similar high-ionization, high-velocity gas in the these three well-studied
Seyferts is an indication that this hitherto undetected component may hold the key to the dynamics
of mass outflow.  

We also analyzed the 2000 HETG spectrum, when the intrinsic continuum flux was $\sim$~2.5 times weaker. 
We find that the heavier X-ray absorption reveals the combined effects of the lower ionization state of
the intervening gas and a higher column density of component D$+$E{\it a}. This requires
$v_{T}$ $\gtrsim$ 1250 km s$^{-1}$. If this is a rotational component, it places the absorber at a 
radial distance of $\lesssim 1.1 x 10^{17}$ cm. Given the predicted $T_{e}$, if X-High is co-located
with D$+$E{\it a}, it cannot be thermally expanding
away (i.e., from the disk). Furthermore, our model constrains X-High to lie well within the point where
its predicted temperature equals its escape temperature. Therefore, the most plausible acceleration mechanism for the
gas is a disk-driven MHD flow. 
  
Although we have been able to probe more deeply than ever into the nature of the intrinsic absorption in NGC 4151, there are still
a number of open issues. Unless the components are co-located, we do not know why the outflow velocities
for X-High and the various components of D$+$E are similar. Determining how the absorbers are radially driven
requires detailed modeling, in particular with MHD codes, that is beyond the scope of this paper. Finally, although we have determined 
that the gas must have a large transverse velocity, we cannot determine whether this is in the plane of the disk 
(i.e., due to rotation) or perpendicular to it.
 
\acknowledgments

S.B.K., I.M.G., and J.R.G. acknowledge support from Smithsonian grant GO2-3138A.
H.N. acknowledges support by the Israel Science Foundation grant 232/03.  E.B. was
supported by grant No. 2002111 from the United States Israel Binational
Foundation. We thank
Gary Ferland for providing us an updated version of Cloudy.

\clearpage

\clearpage

\clearpage


\clearpage

\clearpage

\clearpage

\clearpage

\clearpage

\clearpage

\clearpage

\clearpage

\clearpage

\clearpage

\clearpage
\clearpage

\clearpage

\clearpage
\clearpage

\begin{deluxetable}{lcllll}
\tablecolumns{6}
\footnotesize
\tablecaption{Model Constraints from UV spectra}
\tablewidth{0pt}
\tablehead{
\colhead{Component} & \colhead{$C_{f}^{a}$}  & \colhead{$v_{r}^{b}$} & \colhead{$\sigma_{v}^{c}$} & \colhead{log($N$/cm$^{-2}$)$^{d}$}
& \colhead{Model Prediction}}
\startdata
D$+$E{\it a} & 0.9 & $-491 (\pm 8)$ & 185 & log$N_{OVI}$: $>$ 15.0 & see Table 3\\
             &     & &     & log$N_{NV}$: $>$ 15.0   & \\
	     &     & &     &  log$N_{CIV}$: $>$ 15.0  & \\
D$+$E{\it b} & 0.8 & $-491 (\pm 8)$ & 185 & C~III$^{*}$: log$N_{j=0} < 14.5$ & 14.1$^{e}$\\
& & & &  ~~~~~~~~~~log$N_{j=1} < 14.3$ & 14.3$^{e}$ \\
& & & &  ~~~~~~~~~~log$N_{j=2} \geq 14.8$ & 14.8$^{e}$ \\
& & & &  log$N_{PV}$ $\lesssim 14.0$ & 13.4 \\
D$+$E{\it c} & 0.5 & $-491 (\pm 8)$ & 185 & log$N_{PV}$ $\sim 14.5$ & 14.5\\
E$^{'}$ & 0.8 & $-215 (\pm 11)$ &  26 &  log$N_{SIV}$ $\gtrsim  15.0$ & 15.7\\
& & &  & log$N_{PV}$ $\sim 13.0$ & 13.4\\
\\
\enddata
\tablenotetext{a}{The UV Covering Factor is that of the combined continuum and broad
line emission (see Gabel et al. 2005).}
\tablenotetext{b} {Radial velocity (km s$^{-1}$; Kraemer et al. [2001]).}
\tablenotetext{c}{Velocity dispersion (km s$^{-1}$).}
\tablenotetext{d}{Ionic column densities are indicated as $N_{ion}$, where $ion$ is the 
name of the ion.}
\tablenotetext{e}{Model results for log($n_{H}$) $=$ 8.9.}
\end{deluxetable}
 
\clearpage

\begin{deluxetable}{lrrr}
\tablecolumns{4}
\footnotesize
\tablecaption{Final Model Parameters}
\tablewidth{0pt}
\tablehead{
\colhead{Component} & \colhead{log$U^{a}$} & \colhead{log$N_{H}^{b}$}
& \colhead{$C_{f}$}}
\startdata
X-High & 1.05 & 22.5 & 1.0 \\
D$+$E{\it a} & $-$0.27 & 22.46 & 1.0\\
D$+$E{\it b} & $-$1.67 & 20.8 & 0.9 \\
D$+$E{\it c} & $-$1.08 & 21.6 & 0.6 \\
E$^{'}$ & $-$1.59 & 20.8 & 0.8  \\
X-High-2000$^{c}$ & 0.75 & 22.5 & 1.0 \\
D$+$E{\it a}-2000 & $-$0.57 & 22.93 & 1.0 \\
\\
\enddata
\tablenotetext{a}{The ionization parameters for the screened components include the effects
of absorption by intervening gas.}
\tablenotetext{b}{$N_{H}$ is the total hydrogen column density (in units of cm$^{-2}$), i.e.,
$N_{H}$ $=$ $N_{HI}$ $+$ $N_{HII}$.}
\tablenotetext{c}{X-High-2000 and D$+$E{\it a}-2000 were generated to fit the
2000 HETGS data. Note that the other components of the 2000 model were not included since they are
essentially neutral.}
\end{deluxetable}

\clearpage

\begin{deluxetable}{lccccc}
\tablecolumns{6}
\tiny
\tablecaption{Predicted Column Densities$^{a}$}
\tablewidth{0pt}
\tablehead{
\colhead{Ion} & \colhead{X-High}  & \colhead{D$+$E{\it a}}
& \colhead{D$+$E{\it b}} & \colhead{D$+$E{\it c}}
& \colhead{E$^{'}$}}
\startdata
H~I & 14.0 & 17.0 & 17.1 & 17.4 & 17.3\\
He~II & 15.4 & 19.1 & 19.8 & 16.4 & 19.8\\
C~IV & & 17.2 & 17.1 & 18.0 & 17.1 \\
N~V & & 17.1 & 14.8 & 15.8 & 14.6 \\
O~VI & & 18.5 & 14.1 & 15.8 & \\
O~VII & & 19.1 & & & \\
O~VIII & 17.6 & 18.3 & & & \\
Ne~VIII &  & 17.8 & & & \\
Ne~IX & 15.7 & 17.8 & & & \\
Ne~X & 17.4 & 16.7 & & & \\
Mg~V & & 15.6 & & & \\
Mg~VI & & 16.7 & & & \\
Mg~VII & & 17.3 & & & \\
Mg~VIII & & 17.5 & & & \\
Mg~IX & & 17.4 & & & \\
Mg~X & & 16.9 & & & \\
Mg~XI & 16.0 & 16.3 & & &  \\
Mg~XII & 17.3 & 15.4 & & & \\
\tablebreak
Si~VI & & 16.5 & & & \\
Si~VII & & 17.2 & & & \\
Si~VIII & & 17.5 & & & \\
Si~IX & & 17.4 & & & \\
Si~X & & 17.1 & & & \\
Si~XI & & 16.5 & & & \\
Si~XII & & 15.8 & & & \\
Si~XIII & 16.6 & 15.4 & & & \\
Si~XIV & 17.5 & & & & \\
S~VII & & 16.8 & & & \\
S~VIII & & 17.1 & & & \\
S~IX & & 17.1 & & & \\
S~X & & 16.9 & & & \\
S~XI & & 16.5 & & & \\
S~XII & & 15.9 & & & \\
S~XIII & & 15.0 & & & \\
S~XIV & 15.3 &  & & & \\
S~XV & 16.8 & & & & \\
S~XVI & 17.3 & & & & \\
\tablebreak
Fe~VII & & 15.5 & & & \\
Fe~VIII & & 16.7 & & & \\
Fe~IX & & 17.5 & & & \\
Fe~X & & 17.5 & & & \\
Fe~XI & & 17.4 & & & \\
Fe~XII & & 17.1 & & & \\
Fe~XIII & & 16.6 & & & \\
Fe~XIV  & & 16.1 & & & \\
Fe~XV & 15.7 & 15.5 & & & \\
Fe~XVI & 16.0 & & & & \\
Fe~XVII & 17.3 & & & &\\
Fe~XVIII & 17.7 & & & & \\
Fe~XIX & 17.6 & & & & \\ 
Fe~XX & 17.0 & & & & \\
Fe~XXI & 16.0 & & & & \\
Fe~XXII & 15.0 & & & & \\
\\
\enddata
\tablenotetext{a}{Values are the log of the column densities in units of cm$^{-2}$. 
No value is listed if the
predicted $N_{ion}$ $<$ 14.0 or $<$ 15.0 for those ions constrained by UV or X-ray absorption lines,
respectively. Iron and neon are included for completeness.}
\end{deluxetable}

\clearpage

\end{document}